# Plasmonic Nanocavity to Boost Single Photon Emission from Defects in Thin Hexagonal Boron Nitride

Mohammadjavad Dowran,[1,*] Ufuk Kilic,[2,*] Suvechhya Lamichhane,[3] Adam Erickson,[1] Joshua Barker,[1] Mathias Schubert,[2] Sy-Hwang Liou,[3] Christos Argyropoulos,[4,&] and Abdelghani Laraoui[1,3,&]

[1]*Department of Mechanical & Materials Engineering, University of Nebraska-Lincoln, 900 N 16th Street, W342 NH. Lincoln, NE 68588, Unites States*
[2]*Department of Electrical and Computer Engineering, University of Nebraska-Lincoln, 844 N. 16th Street, Lincoln, NE, 68588, Unites States*
[3]*Department of Physics and Astronomy and the Nebraska Center for Materials and Nanoscience, University of Nebraska-Lincoln, 855 N 16th Street, Lincoln, Nebraska 68588, Unites States*
[4]*Department of Electrical Engineering, The Pennsylvania State University, 203 Electrical Engineering East, University Park, PA, 16802, Unites States*
[*]*Equal contributions*
[&]*Corresponding authors: cfa5361@psu.edu, alaraoui2@unl.edu*

## Abstract

Efficient and compact single photon emission platforms operating at room temperature with ultrafast speed and high brightness will be fundamental components of the emerging quantum communications and computing fields. However, so far, it is very challenging to design practical deterministic single photon emitters based on nanoscale solid-state materials that meet the fast emission rate and strong brightness demands. Here, a solution is provided to this longstanding problem by using metallic nanocavities integrated with hexagonal boron nitride (hBN) flakes with defects acting as nanoscale single photon emitters (SPEs) at room temperature. The presented hybrid nanophotonic structure creates a rapid speedup and large enhancement in single photon emission at room temperature. Hence, the nonclassical light emission performance is substantially improved compared to plain hBN flakes and hBN on gold-layered structures without nanocavity. Extensive theoretical calculations are also performed to accurately model the new hybrid nanophotonic system and prove that the incorporation of plasmonic nanocavity is key to efficient SPE performance. The proposed quantum nanocavity single photon source is expected to be an element of paramount importance to the envisioned room-temperature integrated quantum photonic networks.

## 1. Introduction

The creation of ultrafast, efficient, and bright single photon emission sources are critical components to realize quantum information systems working with photons as qubits. Deterministic single-photon emitters (SPEs) are observed in a wide range of bulk solid-state systems including color centers in diamond such as the nitrogen vacancy (NV)[1–3] and silicon vacancy (SiV) centers,[4,5] defects and divacancies in silicon carbide,[6–9] and semiconductor quantum dots.[10–12] The unique properties of the host semiconductor materials, mainly their high bandgap (*e.g.*, 5.47 eV for diamond), can host nearby nuclear spins with long coherence times ($T_1$, $T_2$) serving as quantum memories/registers.[13,14] These intriguing properties led to the usage of SPEs in many applications including quantum sensing,[15–22] quantum entanglement and teleportation in long distance (> 1 km) based on NV diamond spin qubits,[23,24] and quantum memory enhanced quantum communications that operate at megahertz frequencies using SiV centers in diamond.[5] However, these applications suffer from photon nonradiative losses related to the high internal reflections of bulk substrates (*e.g.*, diamond) and refractive index mismatches when integrated to nanophotonic systems.[3,25] Recently, SPEs have been observed in two-dimensional (2D) materials like tungsten diselenide ($WSe_2$),[26] molybdenum disulfide ($MoS_2$),[27] and hexagonal boron nitride (hBN).[28–35] Ultrathin 2D hosts facilitate enhanced SPE quantum properties compared to bulk systems due to their narrower emission linewidths in the zero-phonon-line (ZPL),[28–30] the ability to inject carriers by applying large electric fields,[36] improved mechanical properties that allow the realization of mechanical oscillators/cavities with high quality factor and high resonance frequency,[37] and ease of integration to optical cavities with extremely-small mode volumes for integrated quantum photonics.[25,38,39]

Despite the extensive recent progress in understanding and utilizing the quantum properties of SPEs in 2D systems such as hBN,[28–35] future developments are severely limited by the difficulty of creating SPEs in hBN nanoflakes with desired spectral properties that match the cavity optical frequency modes,[40] and sufficiently high photon emission rates. Numerous methods have been used to increase the fluorescence rates of SPEs in hBN nanoflakes and films. Atomic force microscopy (AFM) was employed to position gold nanospheres with a diameter of 50 nm in close proximity to SPEs vacancies in hBN flakes, and observed an overall fluorescence enhancement of four times with a radiative quantum efficiency of up to 40%.[41] In our previous work, we spread silver nanocubes (SNCs) on top of hBN flakes with a high density (~ 0.5 SPE/$\mu m^2$) of SPEs, and measured a fluorescence enhancement of 200% and slow emission lifetimes in the order of nanoseconds, limited mainly by the spatial and spectral overlap between SPEs and the weak plasmonic field enhancement caused by bare SNCs.[42] Another approach was used recently by spreading SNCs on top of diamond nanocrystals,[43] CdSe/ZnS quantum dots,[12] and $WSe_2$ flakes[44] deposited on thin metal (Au and Ag) films, hence creating a nanocavity, and an overall 300 – 1900 fold increase in the total fluorescence intensity was obtained. The same metallic nanocavity approach was used very recently to enhance the fluorescence intensity of boron vacancy ($V_B^-$) emitters in hBN flakes up to 480 times, explained by the large Purcell enhancement.[45,46] These findings are promising for using $V_B^-$ and other quantum emitters in hBN in quantum sensing applications.[47,48] However, the aforementioned results were obtained from ensembles of quantum emitters producing classical light that cannot be used in quantum optical communications where the qubits must be single photons. Recently, optical detected magnetic resonance was observed from SPEs in hBN,[49] urging further enhancement of the SPE quantum properties in hBN for quantum sensing applications.[15,18,50] Hence, the metallic nanocavity enhancement in nonclassical light emission from SPEs in hBN is still missing. The creation of this new type of hybrid plasmonic-hBN single photon source will combine a compact ultrathin design, ultrafast nonclassical light emission rate, high brightness, and room temperature operation. All these properties are ideal for designing a novel deterministic single photon source platform to be used in envisioned integrated quantum photonic networks and quantum sensing.

Here, we present, for the first time to our knowledge, an architecture for efficient nonclassical light sources by using metallic nanocavities composed of thin hBN flakes (thickness of 35 nm) sandwiched between small plasmonic SNCs (size ~ 98 nm) and Au thin film (thickness ~ 103 nm) to enhance the quantum properties of SPEs in hBN (Figures 1a and 1b). We demonstrate a substantial enhancement of single photon emission rates at room temperature, manifested by the measured fast emission response in the tens of picoseconds range, due to the designed plasmon nanocavity. Further, we performed rigorous theoretical calculations of excitation enhancement, spontaneous emission enhancement, average electrical field enhancement, and fluorescence enhancement factor of SPEs in SNC/hBN/Au nanocavities. The calculated values fit well the measured ones taking into account the roughness of the Au films. Our results open the door to the usage of SPEs in compact nanophotonic circuits for quantum optical telecommunications[51] and quantum sensing, [15,18,50] in comparison to other systems such as $WSe_2$ that work only at cryogenic temperatures.[44]

## 2. Results and Discussion

### 2.1. Fabrication of Plasmonic Nanocavity

We first exfoliated hBN flakes from hBN bulk crystals (hq graphene) and transferred them to a $SiO_2$/Si substrate (Figure S1a) by using the procedures described in reference [42]. By annealing the hBN/$SiO_2$/Si substrate in the presence of $O_2$ (950 sccm) at a temperature of 1100 °C,[42,52] we created a high density (~ 0.5 SPE/μm$^2$) of stable emitters, confirmed by confocal fluorescence scanning in Figure S1b and anti-bunching $g^{(2)}$ experiments summarized in Table S1 (for selected SPEs) in the Supporting Information (SI) Section S1. The SPEs may originate from oxygen impurities, substitutional oxygen at N or B sites, and $O_2$ etching induced optically active vacancy related defects.[52–54]

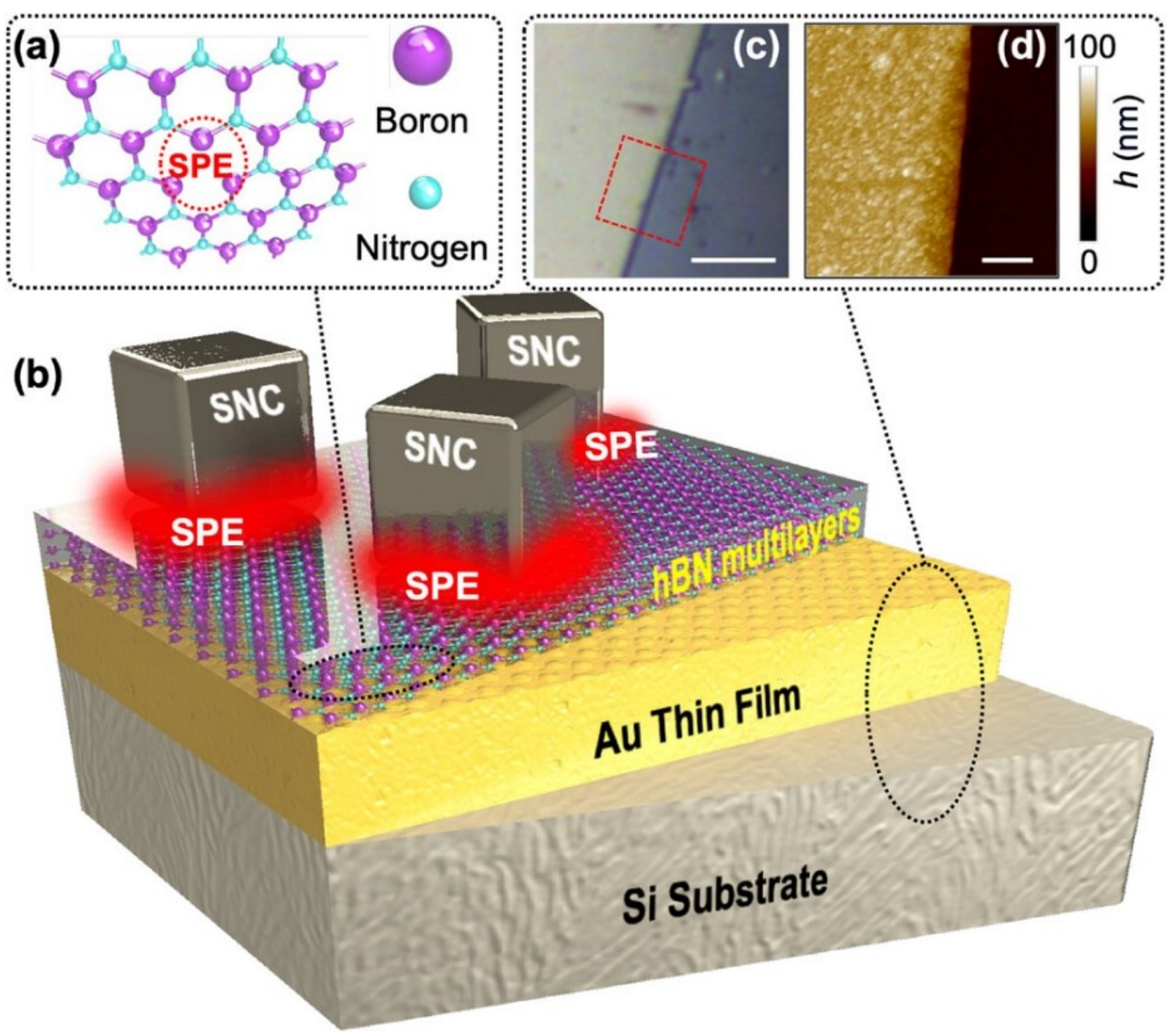

**Figure 1.** (a) Sketch of the hBN flake atoms distribution that exhibits SPE (highlighted with a dashed red circle), composed of nitrogen (purple) and boron (cyan) atoms in a 2D lattice. (b) Schematic of hybrid nanophotonic cavity composed of hBN multilayers sandwiched between the Au film and SNCs to enhance the quantum properties of SPEs. (c) Optical image of the epitaxial grown Au film on a Si (111) substrate. The scale bar in (c) is 5 μm. (d) AFM topography image of the Au/Si substrate highlighted by a dashed square in (c). The scale bar in (d) is 1 μm.

Prior to the transfer of the annealed hBN flakes with SPEs (Figure 1a) to epitaxial gold (Au) films, a Si (111) wafer was cleaned by a sequential sonication process using acetone and isopropyl alcohol to remove any residue. Then, we ion milled the wafer to remove the top Si layer (~ 20 nm thick), to enhance the adhesion during Au deposition. The Au film is sputtered on Si wafer (Figure 1c) at room temperature to achieve a thickness of 103 ± 1 nm, confirmed by AFM in Figure 1d. To perform systematic optical and AFM characterization of the transferred hBN flakes, the Si wafer is marked with a grating of letters. We conducted X-ray diffraction (XRD) analysis of the Au film, revealing a (111) textured gold film (grain size ~ 30 nm), as illustrated in Figure S2a. Following this, we annealed the Au/Si film at 363 °C, resulting in an enhancement of the (111) textured epitaxial characteristics of the Au film, confirmed by a reduction in the XRD linewidth (Figure S2b), and an increase of the Au grains' size to ~ 50 – 150 nm. Further AFM analysis of the Au grains is described in the SI Section S2 and Figure S3.

We followed the procedures described in the SI Section S1 to transfer the hBN flakes with a high density of SPEs from $SiO_2$/Si to Au/Si substrates. It mainly involves using KOH solution to etch down the $SiO_2$ layers[55] under hBN to release the flakes with SPEs to be transferred to Au/Si. While this process is reproducible, it resulted in the etching of the hBN flakes and subsequent reduction of the emitter density to ≤ 0.2 SPE/μm$^2$ (discussed in Section 2.2). After measuring the optical quantum properties of the SPEs in the hBN flakes on top of Au/Si, we spin coated 98 nm silver nanocubes (SNCs) to further enhance the SPE quantum properties due to the plasmonic nanocavity effect. We describe the spin coating process of SNCs on hBN/Au/Si in the SI Section S5.

### 2.2. Optical characterization of SPEs

To measure at room temperature the quantum properties of SPEs in the hBN flakes transferred to Au/Si substrate with and without SNCs, we used a home-built fluorescence microscope described in reference [42]. We first locate the hBN flakes using a CCD camera as shown in Figure 2a and Figure S1c. The thickness of the flake is 35 ± 5 nm, confirmed by AFM (Figure S1f). Then, we perform confocal fluorescence imaging of the selected hBN flake to confirm the presence of bright isolated emitters, spectroscopy to check the SPE emission spectrum, $g^{(2)}$ measurements to confirm single photon emission, and finally lifetime measurements.[42] Figures 2b and S1e display the fluorescence maps of a part of the hBN flake, indicated by dashed squares in Figure 2a and Figure S1c respectively. Figure 2c shows the normalized fluorescence spectrum of a selected SPE (highlighted in Figure 2b) with full width at half-maximum (FWHM) of 5 ± 0.02 nm at wavelength of 639.3 nm, attributed to the ZPL of the defect in the hBN flake.[30] The ZPL peaks of 68 studied SPEs occur in the spectral window of 600 – 760 nm with a negligible phonon side band (PSB), corroborating recent studies performed on SPEs in hBN flakes annealed at high temperature of 1100$^0$C under $O_2$ flow.[42,52]

To confirm the presence of SPEs in the transferred hBN flakes into the Au/Si substrate, we measured anti-bunching $g^{(2)}$ curves in the Hanbury Brown and Twiss (HBT) configuration by collecting the SPE fluorescence using two single photon detection modules (PDM, Micro Photon Devices).[42] We found that > 50% of the emitters in Figure 2b and Figure S1e have a dip of $g^{(2)} <$ 0.4 (see Tables S1 and S2 for some of the SPEs without SNCs).[56] The SPEs are highlighted by dashed and solid squares in Figure 2b and Figure S1e with a density of ~ 0.2 SPE/μm$^2$. We normalized and fitted the short-time scale ($t \lesssim 10$ ns) $g^{(2)}$ response of the SPEs (*e.g.*, emitter SPE with response shown in Figure 2d) to: $g^{(2)}(t) \simeq 1 - (1 + a_1)e^{-t/\tau_1}$, where $a_1$ is the anti-bunching factor and $\tau_1$ is the decay lifetime which includes both the radiative and nonradiative transition lifetimes.[57,58] We obtain $a_1$ = -0.84 ± 0.01 and $\tau_1$ = 1.29 ± 0.02 ns by fitting the curve (scattered filled-circles) in Figure 2d, with $g^{(2)}(0)$ = 0.15 ± 0.01. We further discuss the gold nanograins effects on $g^{(2)}$ in Section 2.3.

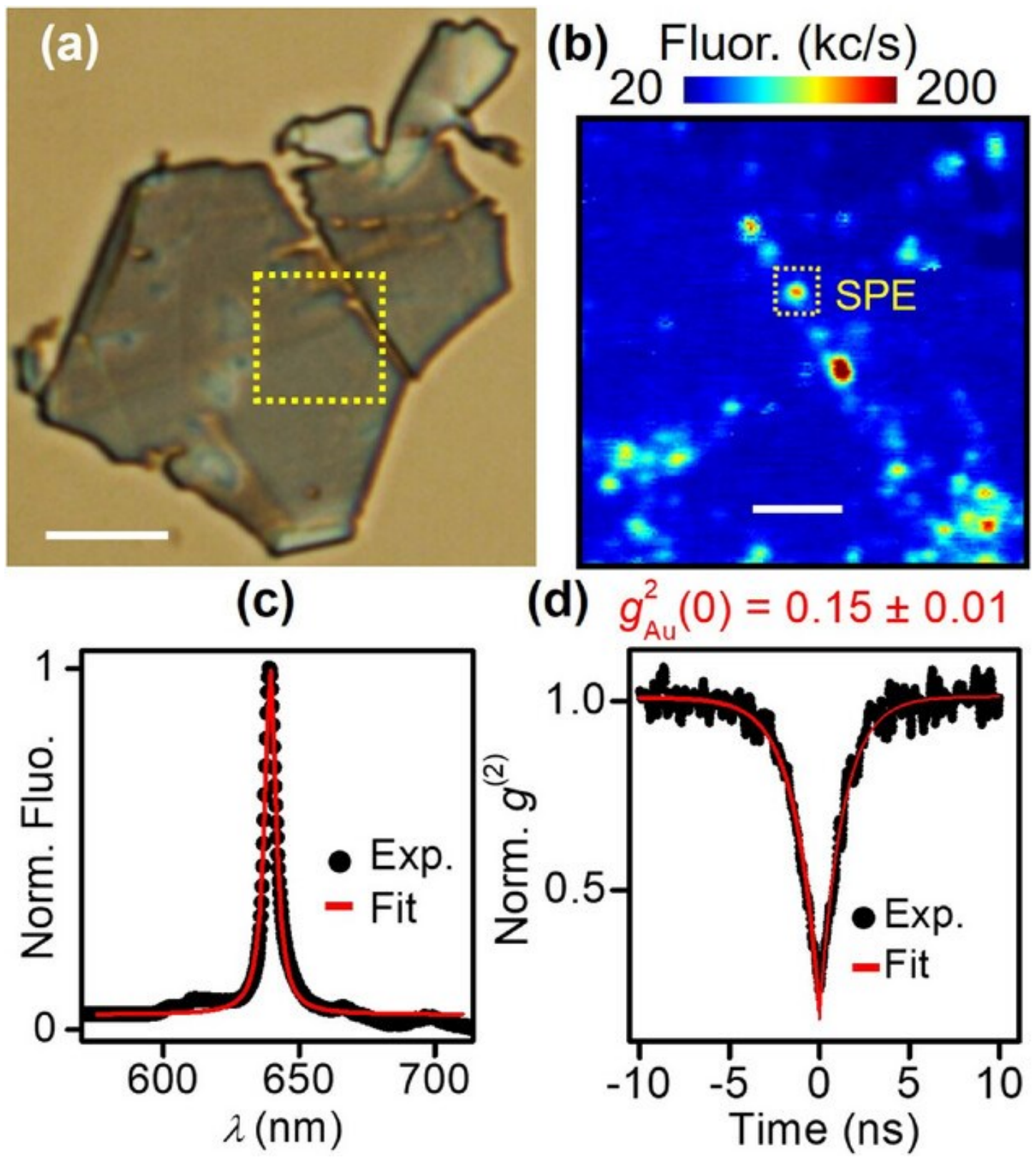

**Figure 2.** (a) Optical image of the studied hBN flake transferred on top of the Au/Si substrate. The scale bar in (a) is 10 µm. (b) Confocal fluorescence map of an hBN flake region highlighted by the dashed square in (a), showing the under-study SPE (inside the square). The scale bar in (b) is 2 µm. (c) Measured (filled circles) fluorescence intensity spectrum as a function of wavelength of a selected SPE marked in (b). The measured curve is fitted (solid line) with a Lorentzian with an FWHM of 5 ± 0.05 nm. (d) Measured (filled circles) autocorrelation $g^{(2)}$ intensity of SPE on Au as function of the delay time between the two single-photon detection modules. The measured curve is fitted (solid line) with $g^{(2)}(t) \simeq 1-(1+a_1)e^{-t/\tau_1}$.

### 2.3. Plasmonic nanocavity enhancement of SPE quantum properties

We measured the quantum properties at room temperature of SPEs in the hBN flake transferred to Au/Si before (Figure 2a) and after spin-coating of 98 nm ± 7 nm SNCs (Figure S7b). To reduce the high intensity autofluorescence (up to 5 × $10^6$ c/s at saturation) of the SNCs in the spectral window 600 – 700 nm,[42] we used a set of bandpass (10 nm) filters to isolate SPEs with a narrow emission spectra. However, some of the autofluorescence is still present in the fluorescence maps after spreading SNC with count rates up to 600 kc/s (see for example Figure S7c). To further identify the location of selected SPEs (highlighted by squares) after spreading SNCs on top of hBN/Au/Si, we used an overlay method and spectral mapping, respectively.[42] Figure 3a shows the fluorescence maps of a selected region of the hBN flake after depositing the SNCs. We can clearly identify most of the SPEs in addition to the SNCs manifested by the small emitting point spread function (PSF) spots inside and outside the flake (see Figure 3a) due to the local confinement of fluorescence.[59] Upon characterizing the hBN flake transferred on Au/Si in comparison to the flakes located on $SiO_2$/Si, we see a clear enhancement of the SPEs quantum properties manifested by an increase of the SPEs average saturation fluorescence intensity $I_\infty$ from ~ 250 kc/s to ~ 1 Mc/s. See for example Figure S1b and Figure S1e measured at a laser power of 1.2 mW. To further assess the effect of Au film and SNCs on the quantum emitters' properties, we measured the second-order photon correlation $g^{(2)}(t)$ of SPEs in the hBN flakes transferred to $SiO_2$/Si, and Au/Si without and with SNCs.

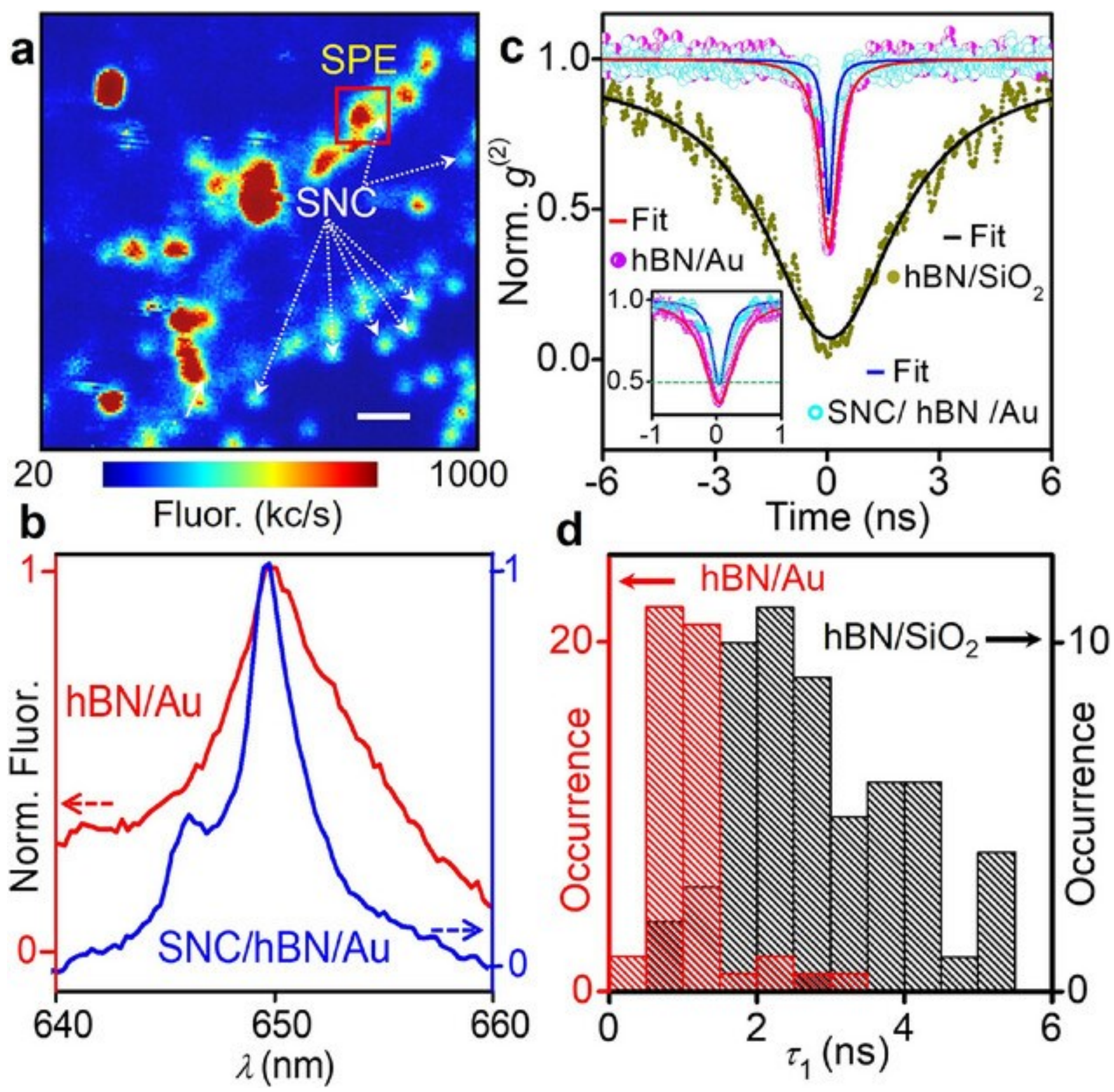

**Figure 3.** ***Plasmonic nanocavity enhancement of SPEs quantum properties in SNC/hBN/Au system***. (a) Fluorescence map of the hBN flake region shown in Figure 2a (indicated by dashed lines) where multiple SNCs are included. The scale bar is 2 μm. The bright emitter SPE (red line box) is coupled with SNC. (b) Measured spectrum of the selected SPE before and after depositing the 98-nm SNCs. (c) Measured autocorrelation $g^{(2)}$ intensity as a function of time for a selected SPE in hBN flake transferred to $SiO_2$ (filled circles) and for SPE (highlighted in a) in the hBN flake transferred to Au/Si substrate without (half-filled circles) and with (open circles) SNCs. In the inset of (c) we plot the zoomed $g^{(2)}$ for SPE on Au with and without SNC, showing $g^{(2)}(0)$ <0.5. The measured curves are fitted (solid lines). (d) Measured $\tau_1$ distribution of 60 SPEs characterized in hBN/$SiO_2$ (black dashed histogram) and 68 SPEs characterized in hBN/Au (red dashed histograms).

We focus here on the enhanced quantum properties of selected SPEs in the hBN flakes deposited on $SiO_2$ (Figure S1b) and Au/Si (SPE inside the solid square in Figure 3a). We see a narrowing of the emission spectrum of the emitter on Au without and with SNC, as clearly depicted in Figure 3b. We observe a significant narrowing of the $g^{(2)}$ response (Figure 3c) with a decrease of $\tau_1$ from 3.1 ± 0.04 ns in hBN/$SiO_2$ to 0.51 ± 0.03 ns in hBN/Au/Si without SNC, and to 0.23 ± 0.04 ns hBN/Au/Si with SNC. While $g^{(2)}(0)$ is 0.013 ± 0.01 for hBN/$SiO_2$, 0.37 ± 0.01 in hBN/Au/Si without SNC, and 0.48 ± 0.01 in hBN/Au/Si with SNC. The increase in $g^{(2)}(0)$ values for SPEs on hBN/Au/Si substrate with SNC is due to the limited time resolution that is constraint by the decay lifetime of the correlation function of our continuous wave (CW) $g^{(2)}$ setup (~ 300 ps).[43] However, all obtained $g^{(2)}(0)$ values are < 0.5 which proves that we indeed see single photon emission in all cases (see Table S1 and Table S2 in the SI). In Table S1 and Table S2 we confirm that most of the SPEs with excited state lifetime $\tau_1 > 0.3$ ns have indeed $g^{(2)}(0) < 0.4$, confirming the single photon emission.[12,42] In Figure 3d, we plot the histograms of $\tau_1$ values extracted from fitting the measured $g^{(2)}$ curves in 60 SPEs in hBN/$SiO_2$ and 68 SPEs in hBN/Au/Si, respectively. We extract a mean $\tau_1$ value of 2.85 ± 1.2 ns in hBN/$SiO_2$ and 0.98 ± 0.33 ns in hBN/Au/Si. The enhancement effect (increase of fluorescence and reduction of $\tau_1$) from just the Au film is explained by the local confinement of electric field at the Au nanograins (See Figure S3d) due to excited plasmonic hotspots,[60,61] which is confirmed by theoretical modeling (see Section 2.4 and Section S7).

To further distinguish between the enhancement effect from the nanostructured Au film and the metallic nanocavity with SNCs, we measured the excited-state lifetime of 60 SPEs in hBN/$SiO_2$ and 68 SPEs in hBN/Au with and without SNCs. A 10 ps pulsed 532-nm laser is used to excite the SPEs and we utilize a time-tagger autocorrelation device (PicoHarp 300) to perform time dependent photon statistics.[42] The time resolution of the lifetime autorotation setup is ~ 19 ps, much finer than CW $g^{(2)}$ setup (300 ps), determined from measuring the instrument response function (IRF, green curve in Figure 4a) with just green laser pulses (see SI section S3 for further details). In Figure 4a, we plot the lifetime curves of SPE in hBN/$SiO_2$, hBN/Au/Si with and without SNCs. The measured curves of SPEs in hBN/$SiO_2$ are well fitted with one exponential decay function $a_3 e^{-t/\tau_{3,\ SiO_2}}$ convoluted with the measured IRF of one of the PDM detectors, where $t$ is time, $\tau_{3,SiO_2}$ is the SPE emission lifetime in hBN/$SiO_2$ that includes both spontaneous and intrinsic nonradiative lifetimes, and $a_3$ is a weighing factor.[30] The $\tau_{3,SiO_2}$ is 5.3 ± 0.6 ns for SPEs in hBN/$SiO_2$ with a mean value of 5.3 ± 1.16 ns, extracted from the distribution histograms of 60 SPEs (see the gray histogram in Figure 4b). The variation of the excited lifetime of SPEs in hBN/$SiO_2$/Si substrate is believed to be related to the location of the emitter $d_{z,SPE}$ and/or the nature of emitter inside the hBN flake.[42]

The measured lifetime curves in hBN/Au without and with SNCs are fitted (solid lines) well with two exponential decay functions convoluted with the measured IRF:[12] $a_{slow} e^{-t/\tau_{slow}} + a_{fast} e^{-t/\tau_{fast}}$, where $a_{slow}$ and $a_{fast}$ are weighing factors. $\tau_{slow}$ is the intrinsic lifetime related to the SPEs in hBN and $\tau_{fast}$ is the spontaneous emission enhancement lifetime related to the environment that includes both the radiative and nonradiative decays.[62] From the measured lifetime curves in Figure 4a on a selected SPE, $\tau_{slow}$ is 0.58 ± 0.02 ns and 0.29 ± 0.01 ns for hBN/Au without and with SNC respectively. Figure 4b shows the distribution histograms of $\tau_{slow}$ from SPEs in hBN/Au without and with SNCs, compared to $\tau_{3,SiO_2}$values obtained from SPEs in hBN/$SiO_2$. We extract mean $\tau_{slow}$ values of 1.15 ± 0.32 ns and 0.88 ± 0.44 ns in hBN/Au without and with SNC simultaneously. However, $\tau_{3,SiO2}$ = 85 ± 2 ps and 25 ± 1 ps in hBN/Au without and with SNC, respectively. The metallic nanocavity substantially speeds up the spontaneous emission lifetime confirmed by measuring $\tau_{fast}$ distribution of 28 SPEs in SNC/hBN/Au nanocavity where enhancement effects are observed with results shown in Figure 4c. We find mean $\tau_{fast}$ ($<\tau_{fast}>$) values of 85 ps and 25 ps for SPEs in hBN/Au without and with SNCs, respectively, meaning approximately more than 3.4 times increase in the spontaneous emission rate when SNCs exist.

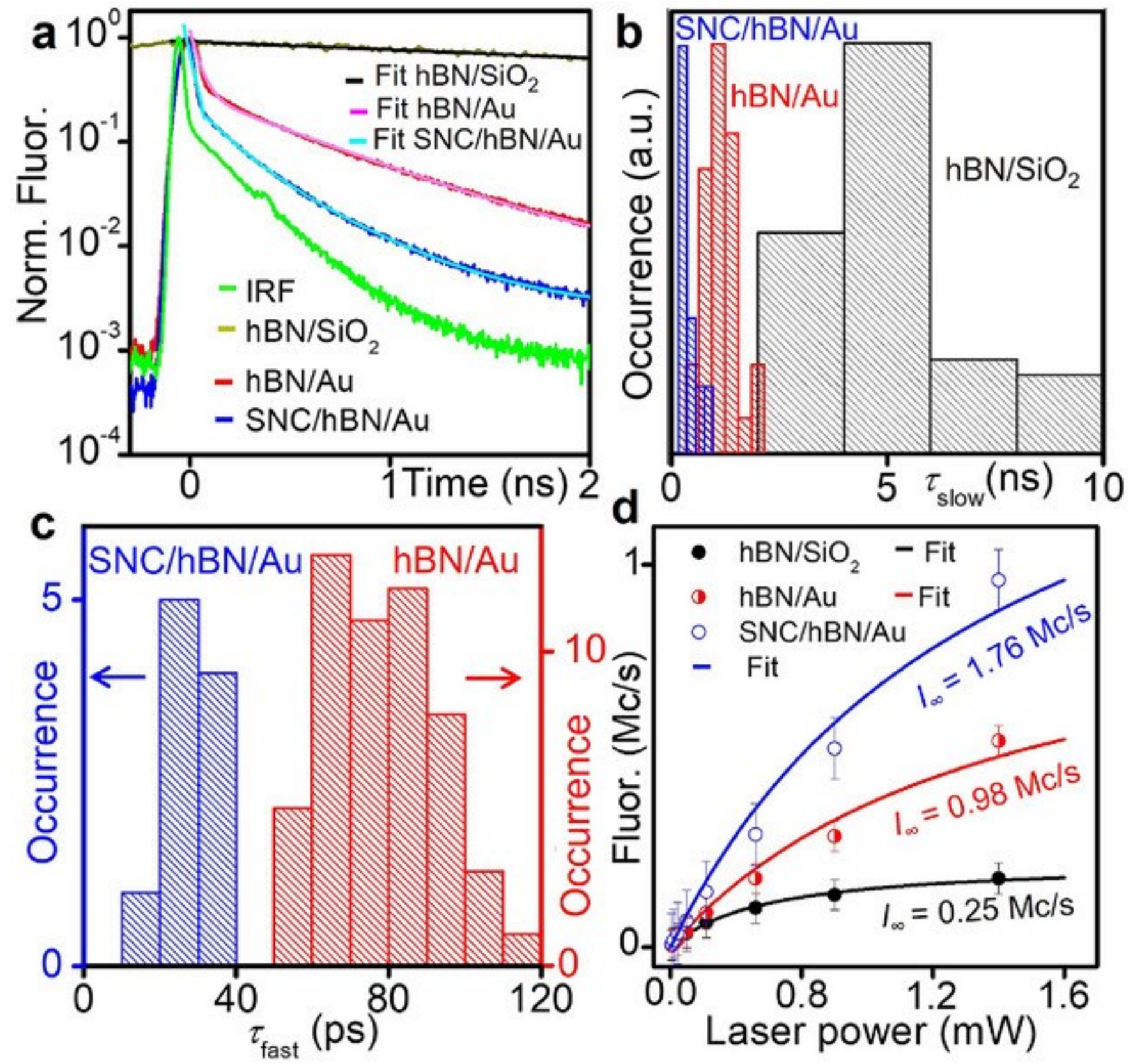

**Figure 4.** ***Plasmon nanocavity-assisted enhancement of SPE quantum properties in hBN***. (a) Measured lifetime response as function of time for a selected SPE in hBN flake transferred to $SiO_2$ (solid dark yellow line) and for SPE in the hBN flake transferred to Au/Si substrate without (solid red line) and with (solid blue line) SNCs. The measured curves are fitted (solid lines). The IRF of the setup is plotted as a solid green line. (b) Measured $\tau_{slow}$ distribution of SPEs characterized in hBN/$SiO_2$ (gray dashed histogram) and in hBN/Au without (red dashed histograms) and with (blue dashed histogram) SNCs. (c) Measured $\tau_{fast}$ distribution of SPEs characterized in hBN/Au without (red dashed histograms) and with (blue dashed histogram) SNCs. (d) Measured (scattered) and calculated (solid lines) fluorescence intensity as function of the CW green laser power for a selected SPE in hBN flake transferred to $SiO_2$ (filled circles) and for SPE (highlighted in Figure 3a) in the hBN flake transferred to Au/Si substrate without (half-filled circles) and with (open circles) SNCs.

The plasmon enhancement of SPE's $\tau_{slow}$ (the intrinsic lifetime) from the Au film in comparison to the $SiO_2$ substrate is: $\frac{1/\langle \tau_{slow\ Au}\rangle}{1/\langle \tau_{3,SiO_2}\rangle} = \frac{\langle \tau_{3,SiO_2}\rangle}{\langle \tau_{slow\ Au}\rangle} = \frac{5.3\ ns}{0.58\ ns} = 9.14$, corresponding to an enhancement factor of ~ 9 times, which is a value comparable to the measured enhancement rates in shallow $V_{B-}$ ensemble emitters in hBN flakes deposited on Au striplines.[61] The introduction of nano-grains/rough surface Au thin film leads to increased local density of optical states due to the Purcell effect, which gives rise to both radiative and nonradiative rates[62–64] depending on the distance of SPEs from the Au grains and their respective sizes (50 – 150 nm). We further detail these effects in the simulation calculations presented in Section 2.4.

We now turn our attention to the effect of SNCs forming a nanocavity on the emission lifetime of the SPEs in hBN/Au. We define the plasmonic nanocavity enhancement of both $\tau_{slow}$ and $\tau_{fast}$ in hBN/Au with and without SNC as: $\frac{1/\langle \tau_{slow\,\mathrm{Au\,w.\,SNC}} \rangle}{1/\langle \tau_{slow\,\mathrm{Au\,w/o\,SNC}} \rangle} = \frac{\langle \tau_{slow\,\mathrm{Au\,w/o\,SNC}} \rangle}{\langle \tau_{slow\,\mathrm{Au\,w.\,SNC}} \rangle} = \frac{0.58\,ns}{0.29\,ns} = 2$, corresponding to an additional enhancement of $\tau_{\mathrm{slow}}$ of 2 with a total enhancement of $\tau_{\mathrm{slow}}$ of 18.28. We also see an enhancement of $\tau_{\mathrm{fast}}$ (spontaneous emission enhancement lifetime): $\frac{1/\langle \tau_{fast\,\mathrm{Au\,w.\,SNC}} \rangle}{1/\langle \tau_{fast\,\mathrm{Au\,w/o\,SNC}} \rangle} = \frac{\langle \tau_{fast\,\mathrm{Au\,w/o\,SNC}} \rangle}{\langle \tau_{fast\,\mathrm{Au\,w.\,SNC}} \rangle} = \frac{85\,ps}{25\,ps} = 3.4$. This is a value very close to the six times increase predicted by simulations later in Figure 6c. The small difference between these two predictions can be attributed to fabrication imperfections that cannot be taken into account in the simulations and the non-ideal dipole alignment of the SPEs that will lead to lower spontaneous emission values in the theoretical calculations. However, all-in-all the agreement between theory and experiment is very good, as discussed later in the paper. It is also commendable that such fast response (25 ps) for SPEs in SNCs has not been reported before experimentally in the literature. This overall enhancement of spontaneous emission lifetime is substantially higher than the slow emission response of hBN without the plasmonic structure, since $\frac{1/\langle \tau_{fast\,\mathrm{Au\,w.\,SNC}} \rangle}{1/\langle \tau_{3,\mathrm{SiO_2}} \rangle} = \frac{\langle \tau_{3,\mathrm{SiO_2}} \rangle}{\langle \tau_{fast\,\mathrm{Au\,w.\,SNC}} \rangle} = \frac{5.3\,ns}{25\,ps} = 212$. Such large enhancement can be explained by the metallic plasmonic nanocavity effect,[51] as observed in other SPE systems such as NVs in nanodiamonds[43] and quantum dots.[12,65] However, our hBN-based nanophotonic structure outperforms these previous studies, since quantum dots suffer from fluorescence blinking[66] and nanodiamonds suffer from NV stability and low quantum efficiency (<< 50%) for smaller nanoscale sizes.[67] In addition, SPEs in hBN have narrow emission lines (< 10 nm in comparison to > 150 nm for NVs), a highly desirable property for quantum photonics applications.[38] The enhancement factor varies across SPEs in SNC/hBN/Au nanocavity system, as summarized in Table S2 (SI Section S6) for selected SPEs. For example, the emitter SPE59 has no enhancement at all and the emitter SPE26 has an enhancement factor of 137.67 times (Figure S12). This effect is explained by either the non-spatial (*e.g.*, SPE59) or non-spectral (*e.g.*, SPE30) overlap between the SPEs and the plasmonic metallic nanocavity modes in the range of 600 – 650 nm and 700 – 750 nm calculated in Figure S10b**.** In some of the SPEs (*e.g.*, SPE26), the plasmonic metal nanocavity modes overlap with the spectral emission wavelength of the emitters leading to narrowing of their ZPL peaks by ~ 50% (see insert of Figure S12b).

However, spontaneous emission enhancement does not mean increased fluorescence, since both radiative and non-radiative rates are enhanced in plasmonic systems.[64] Hence, next, we illustrate the effect of the plasmonic nanocavity in the fluorescence intensity of the SPEs by measuring the saturated count rate $I_\infty$ in hBN/$SiO_2$, and hBN/Au before and after spreading the SNCs (size of 98 nm). In Figure 4d, we plot the measured (scattered curves) fluorescence intensity of SPE in hBN/$SiO_2$ and hBN/Au (without and with SNCs) as a function of the CW green laser power. The saturation curves are fitted (solid lines in Figure 4d) to:[41,42] $I = \frac{I_\infty P}{(P_{sat}+P)}$, where $P_{\mathrm{Sat}}$ is the saturation power and $I_{\infty,SiO2}$ = 0.25 Mc/s, $I_{\infty,hBN/Au}$ = 0.98 Mc/s, and $I_{\infty,SNC/hBN/Au}$ = 1.76 Mc/s for hBN/$SiO_2$, hBN/Au, and SNC/hBN/Au substrates, respectively. The overall enhancement factor of the plasmonic nanocavity is $\frac{I_{\infty,SNC/hBN/Au}}{I_{\infty,SiO2}} = \frac{1.76\,Mc/s}{0.25\,Mc/s} = 7$, corresponding to a fluorescence enhancement of 700%, much higher than the 200% enhancement obtained from just SNC/hBN system reported in reference [42]. We performed similar measurements on 60 SPEs in hBN/$SiO_2$ and 68 SPEs in hBN/Au with and without SNCs and found similar mean fluorescence enhancement rate of 7, see SI Section 6 and Figure S9. Our emission enhancement rates are higher than earlier studies reported on SPEs in hBN integrated with metal nanoparticle array (~2) [68] and non-metallic (*e.g.*, $Si_3N_4$) nanocavities (~ 6).[69] For example in reference [68], the measurements were done on just simple gold plasmonic nanoparticle (NP) array, not nanocavities, *i.e.*, hBN flakes were deposited on Au NPs. The reported fluorescence lifetime enhancement rates in this prior work were ~ 270 – 420 ps, one order of magnitude less than our findings ( ~25 ps). Higher fluorescence enhancement rates (> 10) were reported by using bulk cavity modes (volume ~ $\lambda^3$) [70] and fiber-based Fabry–Pérot (FP) cavities.[71] The bulk cavity increases further the single-photon count rate, even at lower excitation powers, due to an enhanced collection efficiency with the cavity,[70] however, with the drawback of bulk design that will be difficult to be integrated in photonic circuits. Finally, we also compare the emission enhancement with and without SNC when hBN is on top of the Au substrate: $\frac{I_{\infty,SNC/hBN/Au}}{I_{\infty,hBN/Au}} = \frac{1.76\ Mc/s}{0.98\ Mc/s} = 1.8$, which is a close to our theoretical calculations of fluorescence enhancement (*EF* = 2.5) shown in the next section.

## 2.4. Theoretical modeling of plasmonic systems

To gain a comprehensive understanding of the fast and bright single photon emission performance of the proposed nanophotonic systems, we conducted a series of systematic studies based on three-dimensional simulations using the finite element method (FEM). Further details on the simulations created to theoretically investigate the scattering and emission properties, along with additional information on both geometrical and material parameters of the designed single photon emitter are provided in the SI Section S7. More specifically, in Figure S10, we plot the scattering cross sections of both SPE designs with and without SNCs in the spectral range 500 – 750 nm as a function of hBN layer thickness in the range of 25 – 45 nm. Note that the Au substrate is always modelled as a rough surface in both systems to obtain more realistic results. In the case of the absence of SNC, a broad scattering plasmon resonance is obtained which red shifts as a function of hBN layer thickness. On the contrary, when the SNC is present, a narrow resonance is obtained due to the creation of the plasmonic nanocavity, which, interestingly, is also red shifted when the hBN layer thickness is increased.

To understand the measured enhanced quantum properties of the SPE, we illustrate the electric field enhancement factor at the monitoring emission wavelength of 650 nm for SPE designs without and with SNCs, displayed as two-dimensional cross-section slices in Figure 5b and Figure 5e, respectively. The incorporation of SNCs leads to a significant enhancement in the electric field induced inside the hBN layer, with the cross-section field enhancement plots demonstrating up to 5 times increase around the edges of the cube at the specified wavelength.

The spatial distribution of normalized spontaneous emission rate defined as $\gamma_{sp}/\gamma_0$ ($\gamma_{sp}, \gamma_0$ are the spontaneous emission rates of an emitter placed in the presented nanostructures and the same emitter located at free space, respectively) for the hBN/Au structure design without and with SNC are shown in Figure 5c and Figure 5f, respectively. The spontaneous emission becomes stronger at the center of the entire space of the hemi-ellipsoidal inclusions which are used to mimic the roughness of Au thin film (more details are provided in the SI Section S7), leading to the localization of electric field in the proximity of these inclusions. The field gets particularly strong in the location where the hemi-ellipsoidal roughness inclusions approach the hBN layer, and it gradually diminishes as one moves up inside hBN (see Figure 5b). For the SPE emitter design with SNC, we observe that the field is stronger beneath the SNC, but it is primarily concentrated near the bottom corners of the SNC (see Figure 5e) despite the Au thin film being rough also in these simulations. Hence, the coupling of the SPE locations in hBN layer to the plasmonic SNC structure acts as a ultra-small mode volume nanocavity.[72,73]

The incorporation of SNC leads to the creation of a plasmonic nanocavity that efficiently couples to SPEs in the hBN layer and, therefore, increases the photonic local density of optical states which leads to the enhancement in the quantum yield.[74] Hence, the formula to compute quantum yield is QY = $\gamma_r / \gamma_{sp}$ and here is plotted as a function of wavelength (range of 550 – 750 nm, Figure 6e) and emitter depth (range of 2.5 – 30 nm, Figure 6f) for the designs with and without SNC. Interestingly, the QY is improved in the case of the nanocavity (w. SNC) for the entire wavelength range and emitter depths that we currently investigate. This is in direct agreement with the enhanced fluorescence that was measured in the case of the plasmonic nanocavity.

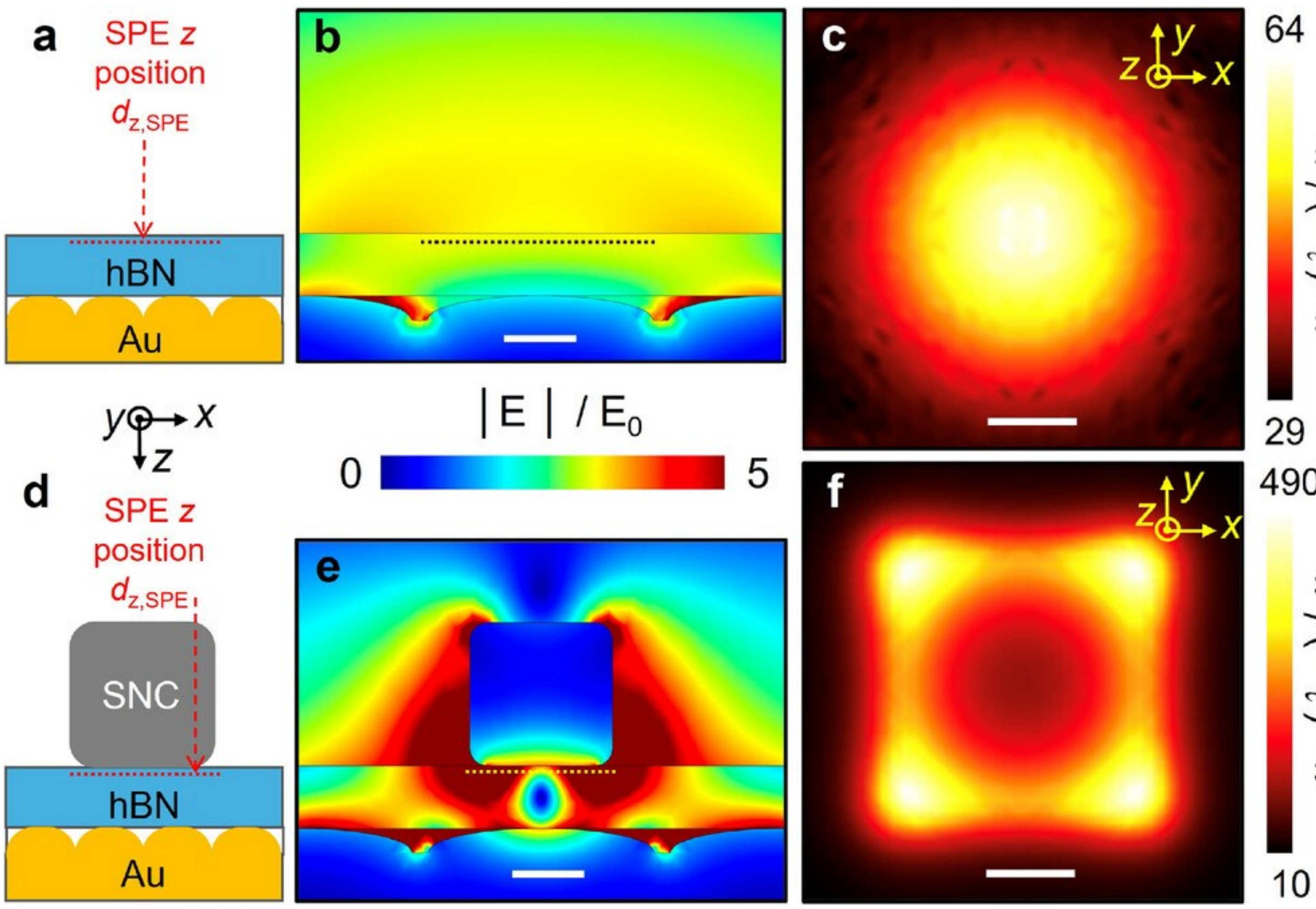

**Figure 5.** ***Simulations of the hybrid SNC/hBN/Au nanophotonic systems***. Schematic illustration from a central cross-sectional viewpoint of SPE designs (a) without and (d) with SNC. The electric field enhancement ($|E|/E_0$) is shown along the central cut-slice of each emitter design without (b) and with (e) SNC at the emission wavelength ($\lambda_{em}$ = 650 nm). The scale bar in (b) and (e) is 40 nm. The spatial distribution of normalized spontaneous emission rate ($\gamma_{sp}(\lambda_{em} = 650\ nm)/\gamma_0$) is displayed in (c) and (f) for the design without and with SNC, respectively. In these plots, the single photon emitter height ($d_{z,\ SPE}$) inside the hBN layer is fixed to 5 nm. The scale bar in (c) and (f) is 20 nm. The emitter height considered in both (c) and (f) is highlighted as dashed line in the schematic illustrations of the designs (a) without and (d) with SNC.

Note that the experimental lifetime results cannot differentiate between radiative and nonradiative contribution to the photon ultrafast emission but provide the total spontaneous emission as was computed in Figures 5c and 5f.[63] Interestingly, based on our simulations, we can differentiate which portion of the total spontaneous emission is radiative and nonradiative, respectively. Hence, to further explore the underlying mechanism responsible for the shortened emission lifetime in the case of plasmonic structures, it is essential to compute the radiative decay rate ($\gamma_r$) and $\gamma_{sp}$ in both structure designs without (Figure 6a) and with (Figure 6b) SNC. Note that the computed $\gamma_{sp}$ is the summation of radiative ($\gamma_r$) plus nonradiative ($\gamma_{nr}$) rates and ideally, we need maximum $\gamma_r$ and minimum $\gamma_{nr}$. In Figure 6c and Figure 6d, the spectral evolutions of the average normalized spontaneous emission ($<\gamma_{sp}\left(d_{\mathrm{z,SPE}}\right)/\gamma_0>$) and radiative decay ($<\gamma_r(d_{\mathrm{z,SPE}})/\gamma_0>$) rates are presented for both plasmonic emitter designs (without and with SNC), respectively, where their obtained maximum values are averaged along the depth $d_{\mathrm{z,SPE}}$ that spans the entire 35 nm thickness of the hBN material. A good agreement is found with measurements of the fast time emission dynamics, which correspond to spontaneous emission enhancement, as stated before in the experimental result section. The computed $\gamma_r$ is close to the total spontaneous emission $\gamma_{sp}$ (that is basically the summation of radiative ($\gamma_r$) plus nonradiative ($\gamma_{nr}$) rates) for both cases without and with SNC at the fixed emission wavelength of 650 nm. Previous studies have already demonstrated that placing the emitter surface close to a metallic thin film leads to quenching and a shortened emission lifetime.[11,63,75–77] Similarly, in the current study, the pronounced roughness of the epitaxial gold thin film leads to some quenching, as can be seen in the lower radiative rate values (Figure 6d) compared to spontaneous emission (Figure 6c), due to the field enhancement (plasmonic hotspots) shown in Figures 5b. and 5e. However, based on our theoretical calculations, the incorporation of the proposed plasmonic nanocavity significantly improves the spontaneous emission performance in the design with SNCs, yielding an average enhancement of approximately six times compared to the design without SNCs at the emission wavelength (650 nm) (see Figure 6c**)**. As discussed in the experimental section, the spontaneous emission rate is inversely proportional to the excited state lifetime, where the experimentally determined lifetime enhancement between these two cases was computed to be 3.4 at the emission wavelength, which is a value very close to the theoretically predicted results.

Finally, the fluorescence enhancement can theoretically be computed by both excitation enhancement ($\gamma_{ex}/\gamma_{ex,0}$) and spontaneous emission ($\gamma_{sp}/\gamma_0$) enhancement, where $\gamma_{ex}/\gamma_{ex,0}$ is directly proportional to the average field enhancement ($|E_{ex}/E_{ex,0}|^2$) inside the hBN SPE computed at the excitation wavelength (532 nm) with ($E_{ex}$) and without ($E_{ex,0}$) the metallic (plasmonic) structures, *i.e.*, computed at free space.[78,79] Hence, we define the fluorescence enhancement factor (*EF*) to theoretically analyze the SPE emission brightness performance of the presented plasmonic systems, given as following:[12,78] $EF=\frac{\eta\,\gamma_{ex}QY}{\eta_0\,\gamma_{ex,0}\,QY_0}$, where the enhancement in quantum yield (QY) is equal to $QY/QY_0$, and the normalized collection efficiency is $\eta/\eta_0$. The latter one represents the probability that an emitted photon will reach the detector and is taken as 0.84 for our case due to our experimental detection approach which is very similar to the one used in reference [12]. The QY of each structure is presented in Figure 6e and Figure 6f. The intrinsic quantum yield of the emitter $QY_0$ is equal to 0.2, as was shown previously for some of the SPEs in hBN.[80] If we substitute all these computed values inside the fluorescence enhancement factor ($EF$) formula and compare the case of without and with SNC, then we get an approximately average value of 2.5 which is very close to the measured fluorescence counts, where we derived that the ratio between the designs with and without SNC is approximately 1.8, as depicted in Figure 4d. Hence, the good agreement between the theoretical enhancement factor predictions and experimental fluorescence counts indicates the accuracy of the simulations in predicting the high radiative quantum efficiency of the emitter design with SNC.[63] Additional factors influencing the overall enhancement of SPE quantum properties in plasmonic nanocavities include the non-spatial and spectral overlap between SNCs (as discussed earlier) and the specific thickness of the hBN film (35 nm). Creating SPEs in thinner (< 15 nm) flakes is expected to substantially increase the fluorescence enhancement factor to > 1000 times, as obtained from SPEs in QDs with a gap of 12 nm.[12]

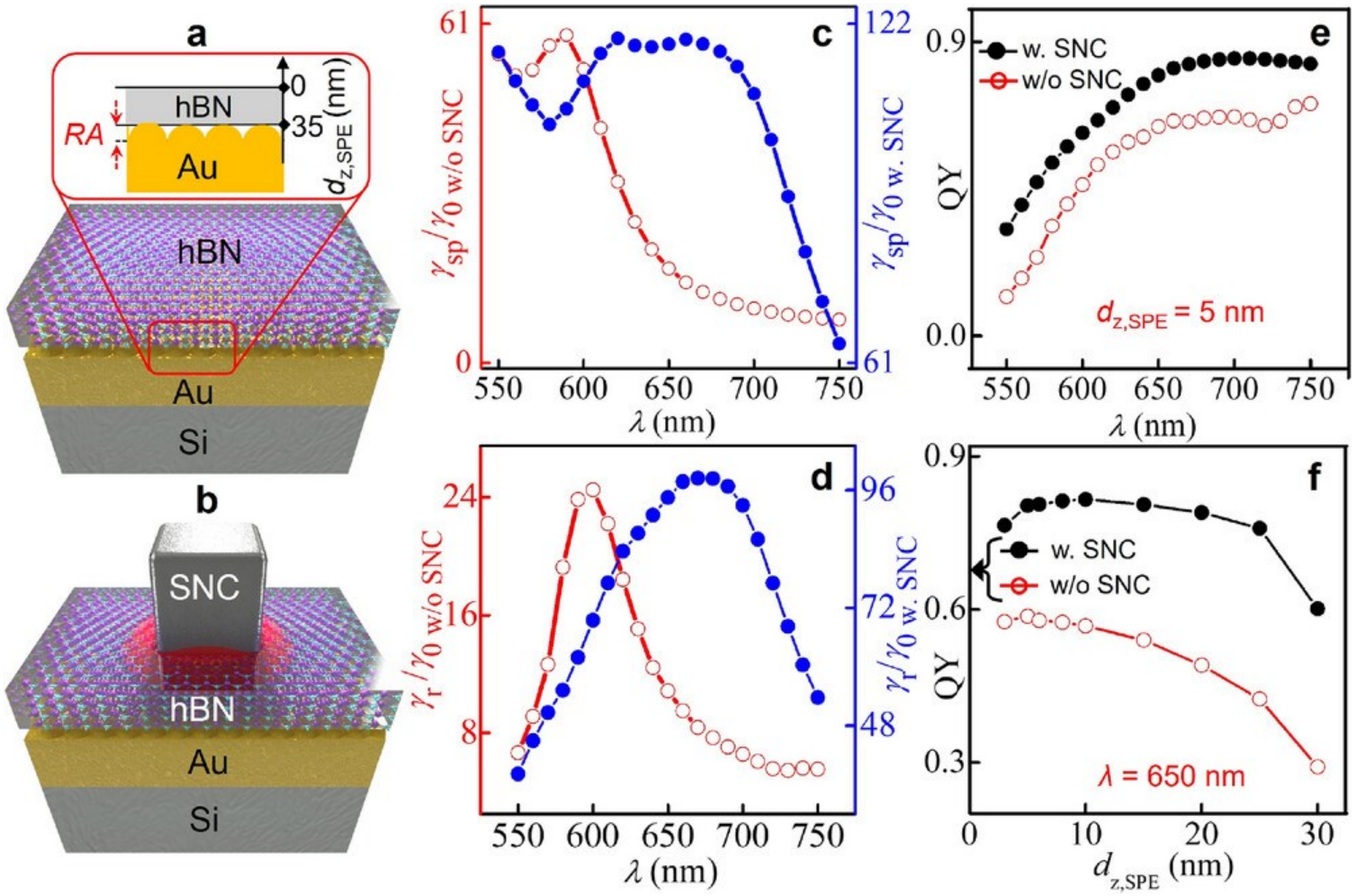

**Figure 6.** ***Spectral and depth theoretical analysis of emission properties*.** (a), (b) SPE emitter designs without and with SNC are schematically illustrated in (a) and (b), respectively. (c), (d) The spectral evolutions of (c) average normalized spontaneous emission rates ($<\gamma_{sp}$ $(d_{z\text{-SPE}})/\gamma_0>$) and (d) average normalized radiative decay rate ($<\gamma_r$ $(d_{z\text{-SPE}})/\gamma_0>$) without (open circles) and with (filled circles) SNCs. (e) The quantum yield spectrum (QY ($d_{z\text{-SPE}}$ = 5 nm)) and (f) its depth profile analysis (QY ($\lambda_{em}$ = 650 nm)) for the emitter design without (open circles) and with (filled circles) SNC.

## 3. Conclusion

In summary, we used metallic nanocavities that sustain gap plasmonic modes, integrated with thin hBN multilayered flakes, to achieve an enhancement in the quantum properties of SPEs at room temperature. We first demonstrated a plasmonic enhancement of the SPEs optical properties from the rough substrate Au nanograins, mainly due to plasmonic hotspots induced along the Au film. To further enhance the quantum properties of SPEs on selected thin hBN flakes, we spin-coated SNCs on the Au/Si substrate. The obtained experimental results are explained by rigorous theoretical modeling based on full-wave simulations, where the roughness in the substrate is also considered. Interestingly, such fast emission response (25 ps) for SPEs in SNCs combined with bright fluorescence properties has not been reported before experimentally in the literature and can have a plethora of quantum photonic applications. For example, the presented plasmonic nanocavity enhanced nonclassical light emission can be useful to increase the optically detected magnetic resonance contrast of SPEs observed recently at room temperature,[49] making hBN a promising material for quantum sensing[15] and spin-based photonic quantum applications.[3] More broadly speaking, our results open the door to the usage of hybrid plasmonic/hBN bright single-photon sources in compact nanophotonic circuits for quantum optical networks that can be used in quantum communications and computing.

**Acknowledgements**

This material is based upon work supported by the NSF EPSCoR RII Track-1: Emergent Quantum Materials and Technologies (EQUATE) Award OIA-2044049. A.L. acknowledges additional support from NSF EPSCoR Award OIA-2429381. The research was performed in part in the Nebraska Nanoscale Facility: National Nanotechnology Coordinated Infrastructure and the Nebraska Center for Materials and Nanoscience (and/or NERCF), which are supported by NSF ECCS under Award 2025298, and the Nebraska Research Initiative. C.A. acknowledges partial support from NSF DMR Award 2224456. M.S. acknowledges partial support from NSF ECCS Award 2329940 and AFOSR Award FA9550-21-1-0259.

**Conflict of Interest**

The authors declare no competing financial interest.

**Author Contributions**

M.D. and U.K. contributed equally to this work and are co-first authors. M.D. synthesized the hBN flakes, performed the optical measurements and analyzed the data. U.K. performed theoretical modeling, AFM, and Ellipsometry measurements. S.L and S.-H.L prepared the $SiO_2$/Si and Au/Si substrates and made the marks. A.E. assisted M.D. in the AFM measurements. J.B. performed fluorescence polarization dependence of selected SPEs on $SiO_2$. C.A. and M.S. assisted U.K. in the theoretical calculations. A.L. and C.A. designed the experiments and supervised the project. A.L., C.A., and U.K. wrote the manuscript with contributions of all authors. All authors have given approval to the final version of the manuscript.

**Data Availability Statement**

The data that support the findings of this study are available from the corresponding author upon reasonable request.

## Supporting Information

### S1. Exfoliating the hBN flakes on Si/$SiO_2$ and Au/Si substrates

$SiO_2$/Si and Si (111) coated with Au film were marked with reference grid marks (letters) by using laser lithography (DWL-66). The marked substrates were then sonicated in acetone to clean them from mask residue, followed by plasma processing at 80 W under 15 sccm of oxygen flow for 10 minutes. The substrates have maximum acceptance of the flakes within 10 minutes after plasma processing. For exfoliating hBN flakes on the $SiO_2$/Si substrate from bulk hBN crystals (hq graphene), we followed the recipe described in reference [42]. After annealing the hBN/$SiO_2$/Si substrate at high temperature (1100 °C) for 4 hours under $O_2$ flow (950 sccm) we used confocal fluorescence microscopy and $g^{(2)}$ antibunching experiments to confirm the presence of high density (> 0.5 SPE/1μ$m^2$) of stable single photon emitters (SPEs).[42] We spin coated first polyvinyl alcohol (PVA) and then polymethyl methacrylate (PMMA) at a speed of 3000 rpm for 60 seconds respectively on top of hBN/$SiO_2$/Si. After this, we baked the solution at 120 °C for 20 minutes on top of a hot plate. We then dipped the PMMA/PVA/hBN/$SiO_2$/Si substrate into 2 mole potassium hydroxide (KOH, density of 56.1 g/mol) solution in a glass beaker. We then heated the beaker at a temperature in the range of 100 – 120 °C for 2 hours until bubbles form and the PMMA/PVA/hBN membranes float. To detach the PMMA/PVA/hBN from the $SiO_2$/Si we shook it occasionally. We grabbed the membrane using a flat metal/glass piece and dipped it in another beaker with hot Di water to remove any KOH residue. To prepare the Au/Si for the new transfer of hBN flakes with SPEs, we dipped it into another beaker full of water and added the PMMA/PVA/hBN membrane. We carefully checked the orientation of the PMMA/PVA/hBN membrane on top of Au/Si substrate under the optical microscope by looking at the letters on the marked grid, i.e., the letters should not be mirrored and should be readable correctly. The PMMA/PVA/hBN/Au/Si is dried up by heating it on a hot plate at 40 °C for 10 minutes, then at 110 °C for 15 minutes. To remove PMMA and PVA from hBN/Au/Si substrate we dipped it in acetone for 20 minutes and then in DI water for 20 minutes respectively. We finally cleaned the hBN/Au/Si substrate by using plasma processing at a power of 60 W for minutes under $O_2$ flow (15 sccm). The later step is repeated many times to remove any PMMA and PVA residues without etching out the hBN flakes.

We performed optical and topography measurements at room temperature of the hBN flakes on $SiO_2$/Si before and after transferring to Au/Si (111) substrates, summarized in Figure S1. With an optical microscope, we check the distribution and thickness of the hBN flakes transferred to $SiO_2$/Si and Au/Si substrates. hBN flakes of different thicknesses transferred to $SiO_2$/Si substrate show distinct colors depending on their thicknesses (Figure S1a), which were previously characterized using atomic force microscopy (AFM) analysis. We measured the hBN flakes (insert of Figure S1a and Figure S1d) thickness by using AFM (Innova from Bruker) and related it to their color under the optical microscope. We used AFM tips Tap300-G (Budget sensors) in the tapping mode with a force constant of 40 N/m and resonance frequency of ~ 300 kHz. In the inset of Figure S1a we display the lateral cross line of the hBN flake transferred into $SiO_2$/Si substrate (highlighted by dashed square) and measured a height of 24 nm. Figure S1b shows the fluorescence intensity map of the hBN flake, highlighted by dashed square in Figure S1a, having at least 26 SPEs labeled with solid and dashed squares.

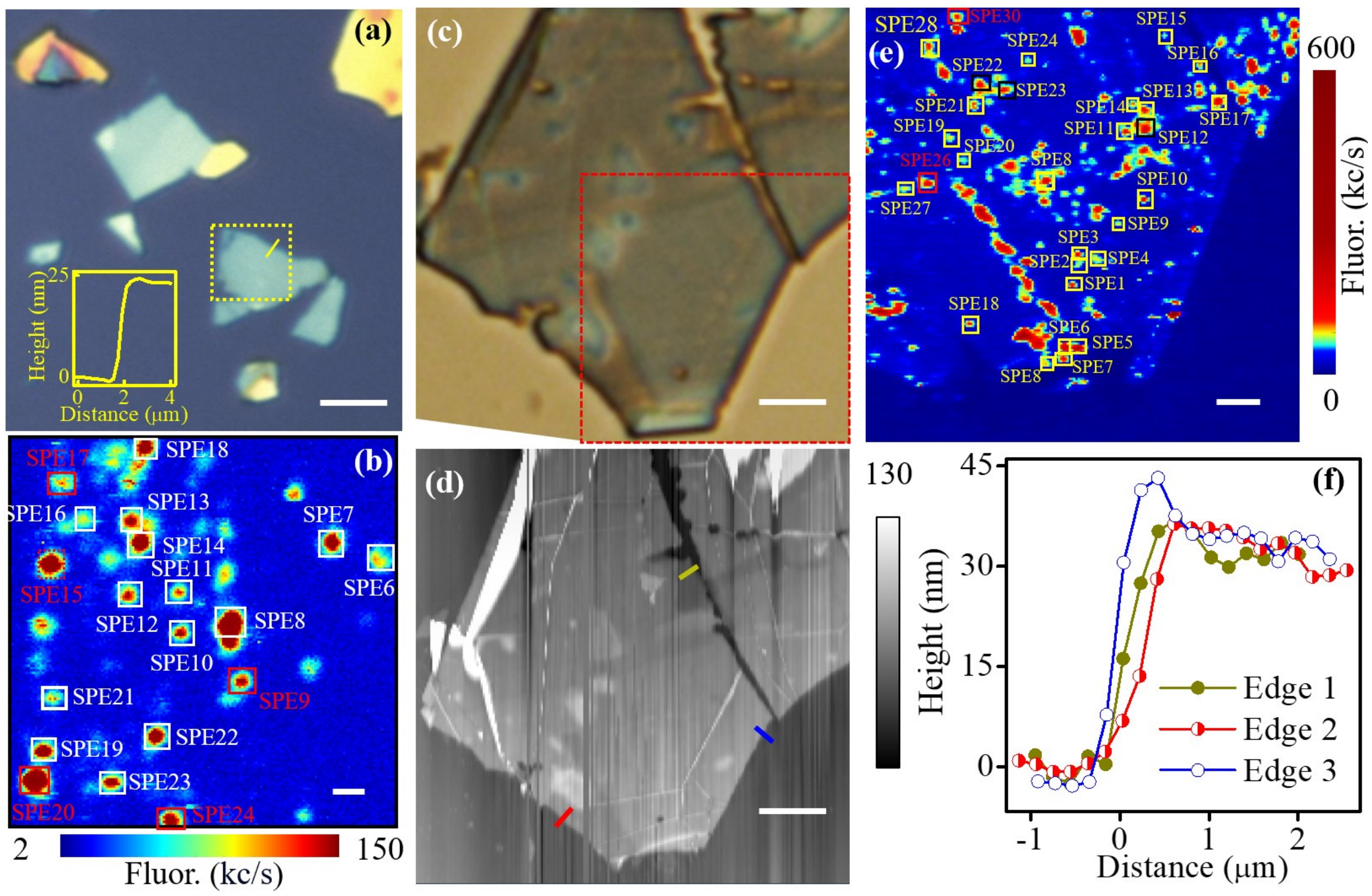

**Figure S1.** (a) Optical image of some of the hBN flakes transferred to $SiO_2$/Si substrate. The scale bar is 10 μm. Lower insert of (a): Measured AFM lateral cross section (yellow solid line) of height profile across the hBN flake in the dashed square. (b) Fluorescence intensity map of the region highlighted by a dashed yellow square in (a). $g^{(2)}$ measurements confirm that most of the emitters are SPEs, highlighted with solid and dashed squares in (b). The scale bar is 1 μm. Optical (c) and AFM (d) image of part of hBN flake studied in the main text, transferred to Au/Si (111) substrate. The scale bar is (c) and (d) is 5 μm. (e) Fluorescence intensity map of the region highlighted by a dashed red square in (c). The scale bar is 1 μm. (f) The height of the measured lateral cross sections (solid lines) from the AFM image in (d) shows the thickness of the flake is uniform ~ 35 ± 5 nm.

We then characterized the optical and topography of the hBN flakes transferred to Au/Si (111) substrates. Figure S1c and Figure S1d display the optical and topography AFM image of part of the hBN flake studied in the main text (Figure 2a) respectively. We measured a uniform thickness of the hBN flake of 35 ± 5 nm, confirmed by the height profile of the lateral cross sections in Figure S1f. Figure S1e exhibits the fluorescence intensity map of part of the hBN flake, highlighted by a dashed red square in Figure S1c. A clear enhancement of the SPEs fluorescence intensity in the hBN flake on Au/Si (up to 600 kc/s) in comparison to the hBN flake on $SiO_2$/Si (up to 150 kc/s). This is explained by the local plasmonic hotspots formed in the gold (Au) nano-grains.[61]

## S2. Growth of epitaxial gold films

We used ion beam deposition to sputter a 4-inch Si (111) wafer with a layer of Au. The Si wafer underwent a sequential sonication process using acetone and isopropyl alcohol to clean it from any residue. We then ion milled the wafer to remove the top Si layer, approximately 20 nm thick, to enhance adhesion prior to Au deposition. The gold is sputtered at room temperature, with a deposition rate of ~ 7 nm per minute, for a total duration of 14 minutes, resulting in ~ 103 nm thick film, confirmed by AFM measurements (discussed below). After the grating fabrication, we conducted X-ray diffraction (XRD) analysis of the Au films, revealing a (111) textured gold film, as illustrated in Figure S2a. Following this, we subjected the Au film to annealing near the eutectic temperature of 363 °C of Au/Si, in a vacuum for approximately 30 minutes. This annealing process notably enhanced the (111) textured characteristics of the Au film and resulted in an observable reduction in line width (Figure S2b), effectively indicating an increase in grain size, confirmed by AFM measurements in Figure S3.

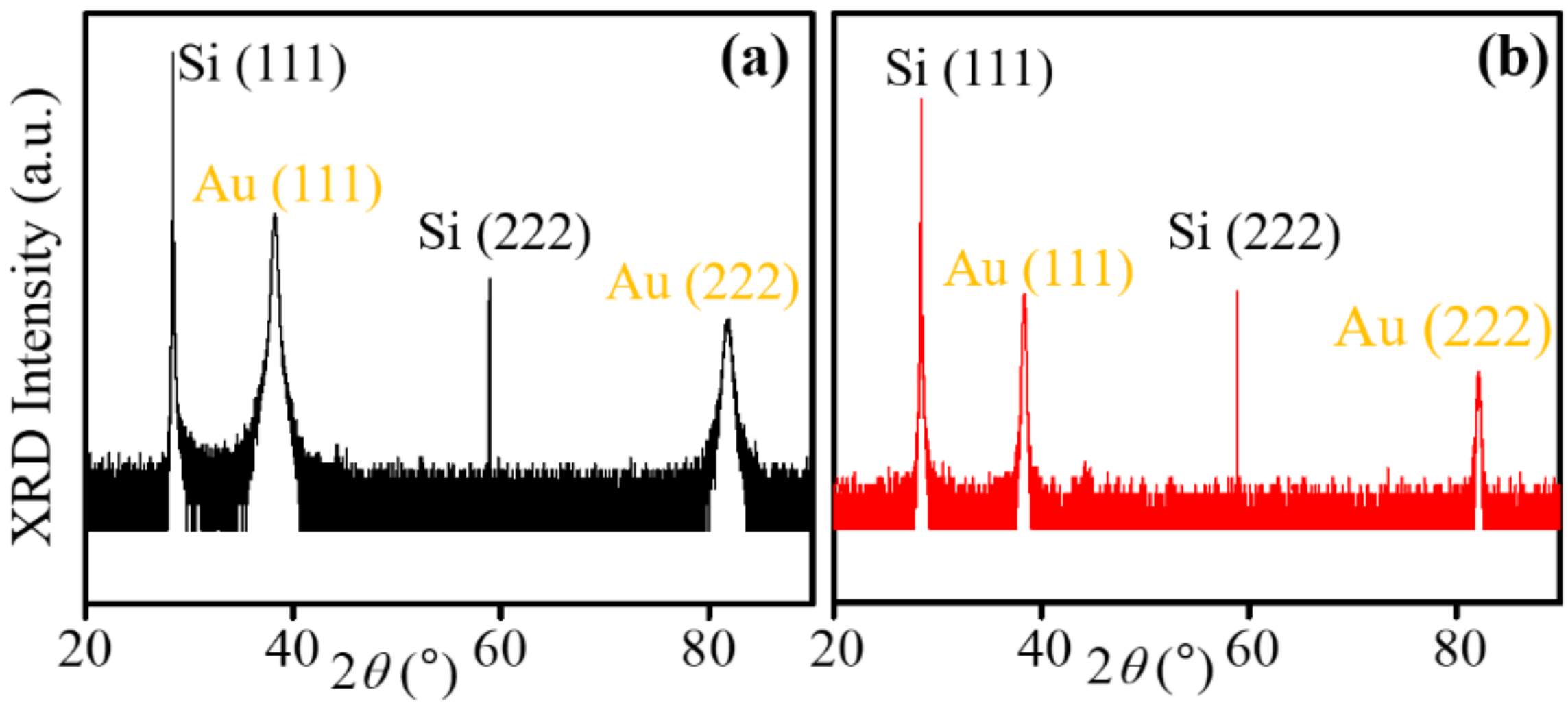

**Figure S2.** XRD spectrum for Au sputtered on 4-inch Si (111) wafer before (a) and after annealing (b). The Au peaks are narrowed upon annealing, explained by the increase of grain size.

This phenomenon is further demonstrated by AFM imaging on the Au film (see optical image in Figure S3a), both before and after annealing. The AFM image of the gold film before and after (Figure S3b) annealing depicts the (111) increase in textured grains of size from ~ 90 – 200 nm. From the ellipsometry analysis (not reported here), we found that Au layer thickness is found to be optically thick enough which means the substrate reflections from the Au/Si wafer interface do not contribute to the acquired ellipsometry data. AFM image, which is shown in Figure S3b, is obtained by using the Bruker Dimension Icon AFM with automatic image optimization mode based on Peak-Force tapping technology. The field of view was chosen to be 5 × 5 μm$^2$ with a line resolution of 512 × 512 with a scan velocity of 0.1 lines per second. Nanoscope Visualization and Analysis software is utilized to perform detailed roughness and grain size analysis. The average ($R_a$) and root mean-square ($R_q$) values of surface roughness were obtained as 7.16 nm and 5.65 nm, respectively. Grain size analysis shows the presence of two major accumulations of grain size values which are clearly seen from the histogram plot (see Figures S3c). As is seen from Figure S3d, the sample also has the region (dark section) without Au thin film that enables the extraction of thin film thickness. The average Au thickness is about 103 ± 1 nm.

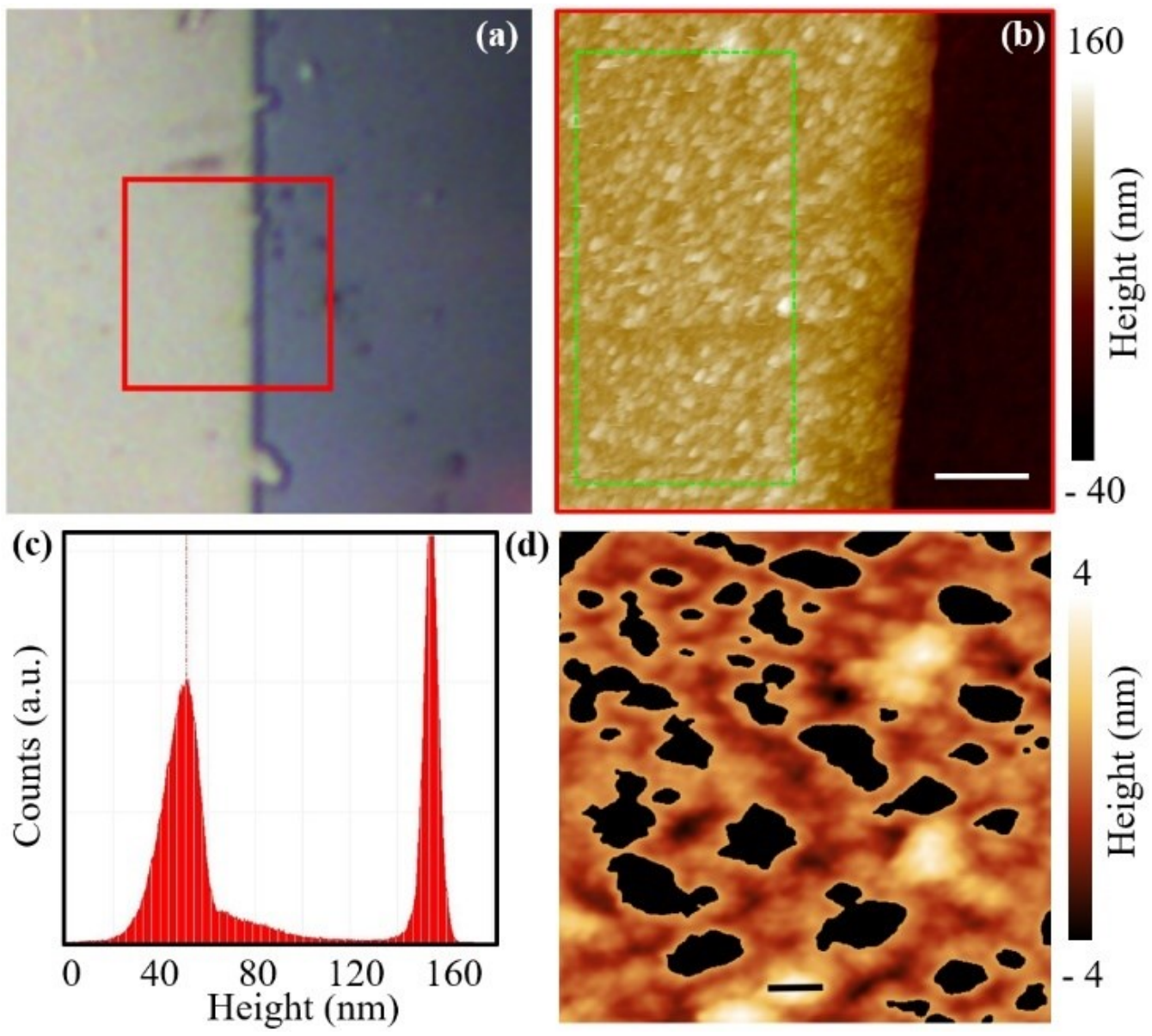

**Figure S3.** (a) Optical image of the Au/Si substrate. (b) AFM image of Au/Si substrate showing a thickness of the Au film to be ~ 100 nm. (c) Measured size distribution of the region highlighted by a dashed green square in (b), confirming two size distributions centered at 50 nm and 150 nm respectively. (d) Dark mask AFM analysis confirming the presence of the two sizes (~ 50 nm and ~ 150 nm) of Au grains.

## S3. Time domain instrument response function (IRF) of the experimental setup

To measure the excited and emission lifetimes of the emitters in the hBN flake, we used a 10 ps pulsed green laser (SuperK Fianium FIU-15 from NKT Photonics, repetition rate of 78 MHz) to excite the emitters and a single photon detecting avalanche photodiode (MPD, PDM Series 50 µm) to collect the fluorescence.[81] The output of the PDM is connected to a timing module with a resolution of 4 ps (PicoQuant PicoHarp 300), which detects the arrival time of each photon. This technique, known as time-correlated single photon counting, results in a histogram of photon arrival times[81] which corresponds to the time-dependent rate of photon emission from the emitter. The trigger for the timing module is created by splitting part of green 10 ps laser beam before the objective/dichroic mirror and connecting it to the timing module Picoharp 300. The overall timing resolution of the system is ~19 ps, as determined by measuring the reflected laser light from the substrate, see Figure S4. The time jitter of the PDM module is < 15 ps. The curve is fitted with a Gaussian function with duration (full width at half-maximum, FWHM) of 19 ps and an exponential decay for the tail (113 ps). We included the fitting of SPE in Figure 4a (main text), performed by convolving the measured IRF with a biexponential function following reference [12], see Figures S4b and S4c.

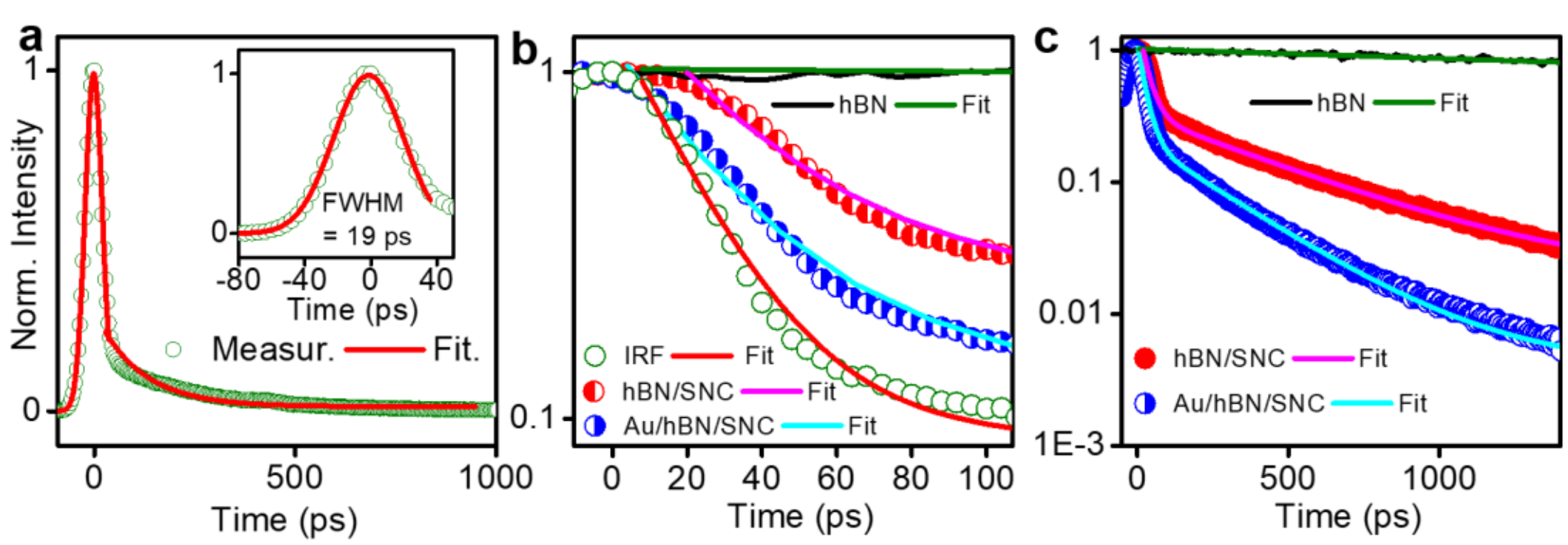

**Figure S4.** IRF of the lifetime resolved optical setup. The solid lines in (a) and in the inset of (a) are fits of the measured IRF, leading to a resolution of ~19 ps of our setup. Measured and fitted lifetimes of the SPE studied in Figure 4 at shorter (b) and longer (c) times.

## S4. Optical and quantum properties of SPEs created on $SiO_2$/Si substrates

In Table S1 we list the optical properties at room temperature (fluorescence spectrum, $g^{(2)}$ curve, lifetime curve, and fluorescence intensity vs excitation laser power saturation curve) of five selected SPEs from hBN/$SiO_2$ inside the solid red squares in Figure S1b, labeled with the numbers SPE9, SPE15, SPE17, SPE20, and SPE24. Most of the emitters in Figure S1b are SPEs with a dip of $g^{(2)} < 0.25$.

**Table S1**

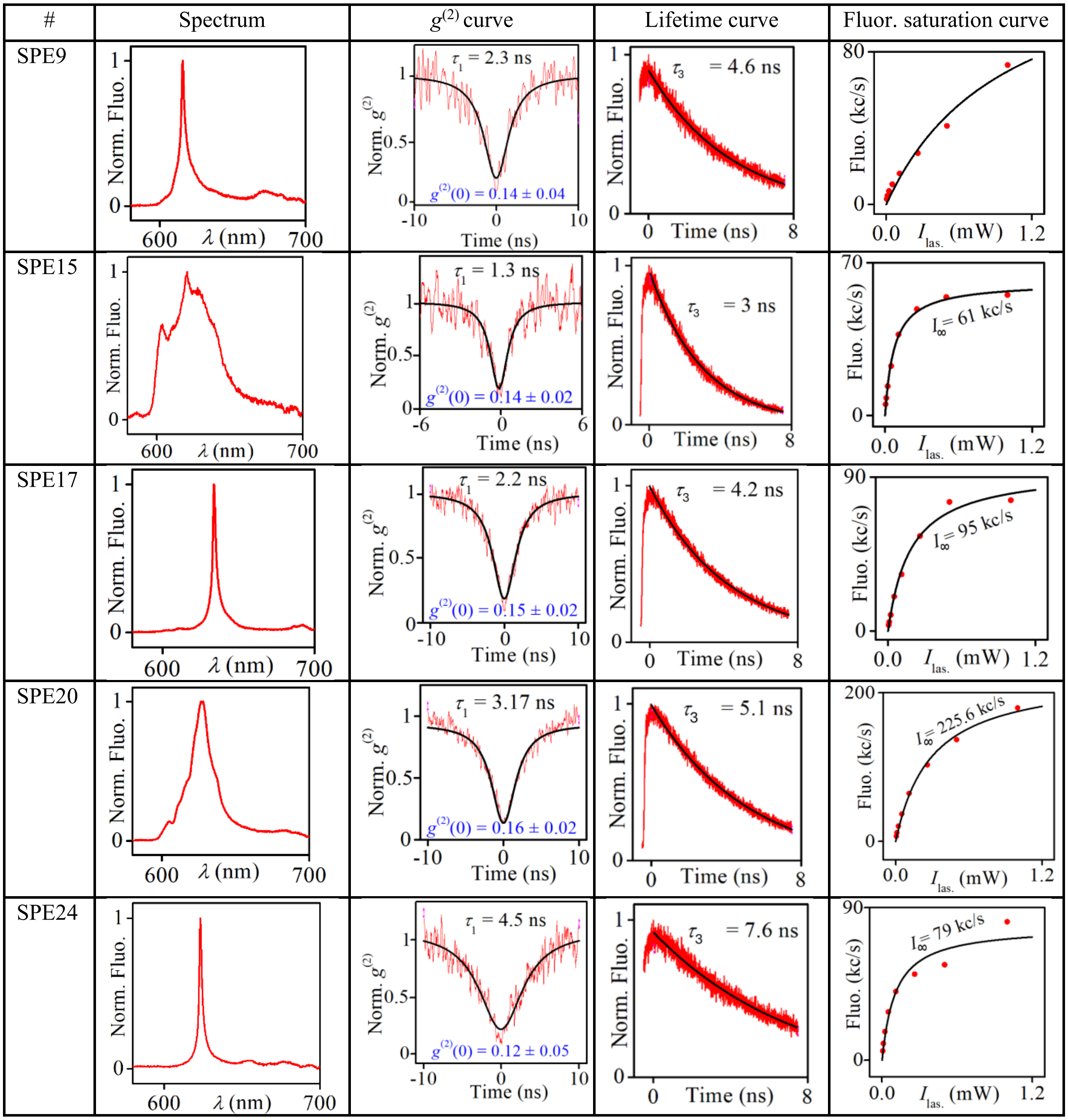

In Figure S5, we plot the histogram of the $g^{(2)}(0)$ values of the 60 SPEs studied in the hBN flake transferred to $SiO_2/Si$ substrate in Figures S1a and S1b. Indeed, ~84% of the SPEs have $g^{(2)}(0)$ dip < 0.2.

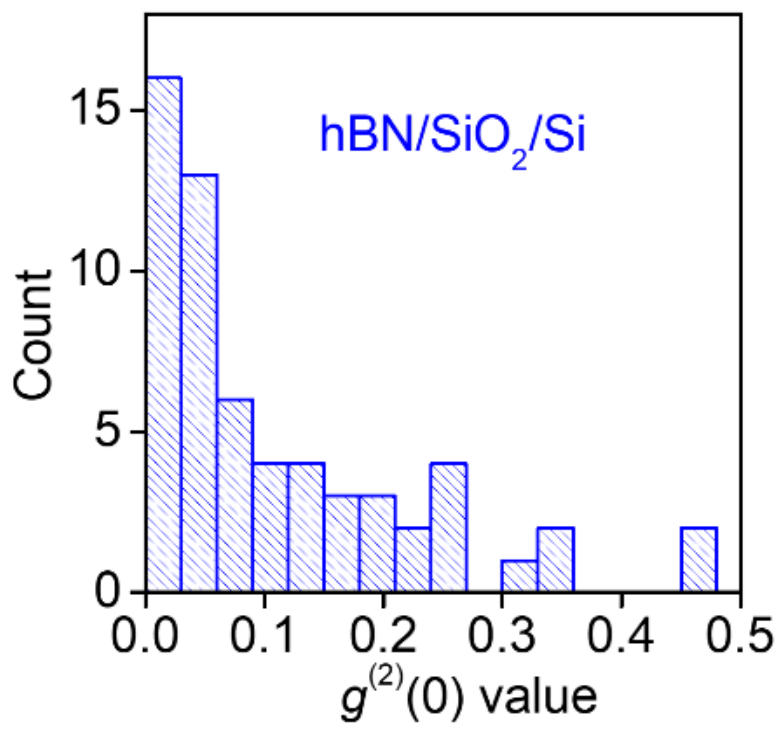

**Figure S5.** Histogram of $g^{(2)}(0)$ values of (a) 60 SPEs studied on $hBN/SiO_2/Si$

To check the orientation of the SPEs optical dipoles, we performed fluorescence polarization dependence of a selected emitters SPE1 (Figure S6a) and SPE2 (Figure S5c) with $g^{(2)}(0) = 0.05 \pm 0.01$ and $0.24 \pm 0.01$, respectively. Figures S6b and S6e show the emission polarization data of SPE1 and SPE2, respectively. The fluorescence changes slightly with the 532 nm laser polarization, meaning that the dipoles of SPEs are randomly distributed inside the hBN flake. This is different from earlier measurements done on SPEs in bulk hBN crystals with single dipole transition (linearly polarized along the basal plane of hBN).[82] The overall change of the fluorescence intensity with the polarization angle is ~ 32% of SPE1 (Figure S6c) and ~ 17% for SPE2 (Figure S6c). These variations are less than the fluorescence enhancement (> 350%) obtained from the Au grains and the metallic nanocavity Au/hBN/silver nanocube, as discussed in the main manuscript and in Section S6.

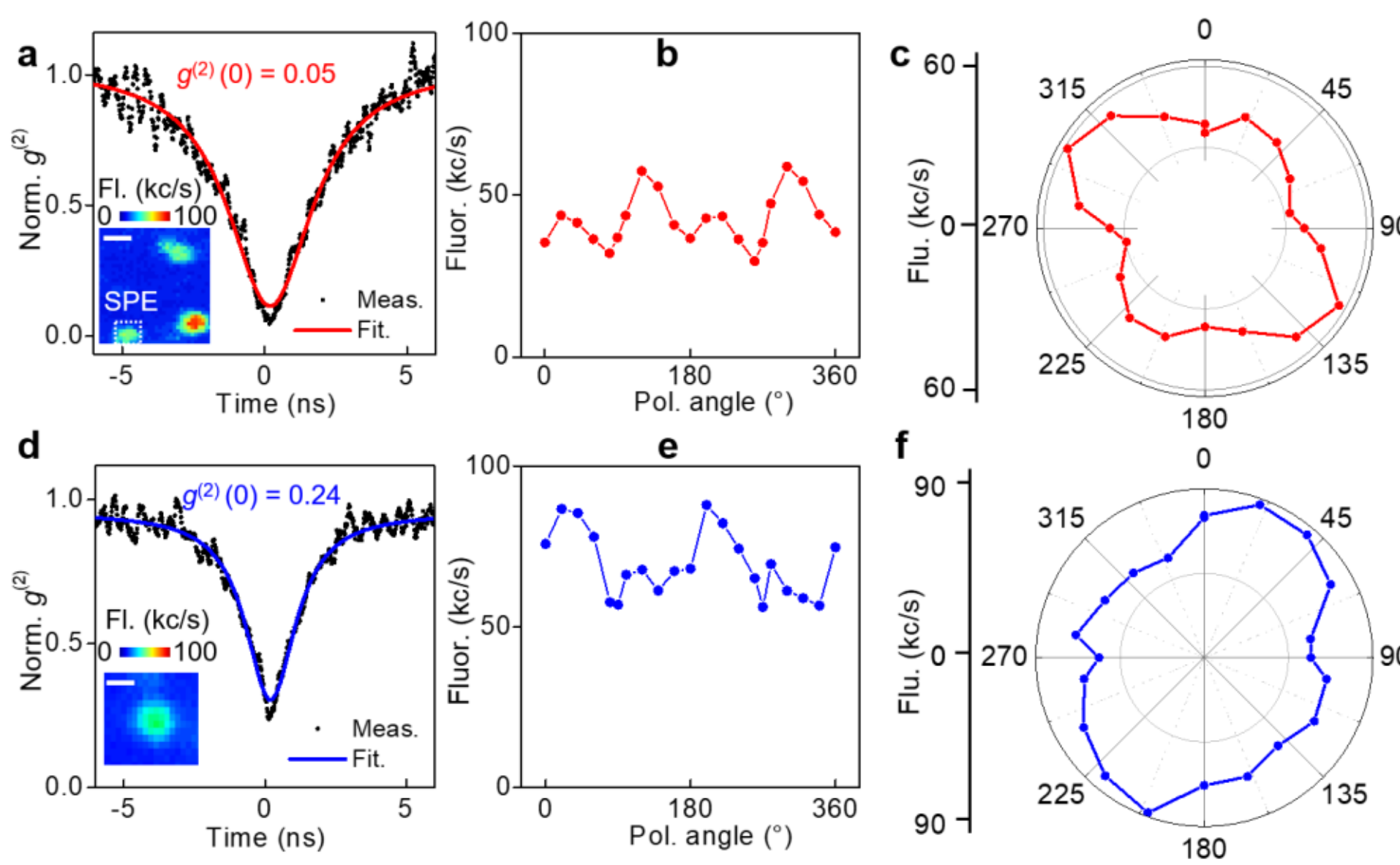

**Figure S6.** Measured (filled circles) autocorrelation $g^{(2)}$ intensity as a function of time for (a) SPE1 and (d) SP2 in hBN flake transferred to $SiO_2/Si$ and fitted (solid lines) to a Lorentzian function. The insets of (a) and (d) are the fluorescence maps of the emitters SPE1 and SPE2, respectively (the scale bar is 0.5 µm). Polar fluorescence intensity map as a function of the angle of a polarizer placed before the objective for (b) SPE1 and (e) SPE2. Fluorescence intensity versus the angle of a polarizer showing ~ 32% and ~ 17% deviations for SPE1(c) and SPE2 (f).

## S5. SPEs characterization before and after spreading of silver nanocubes

Silver nanocubes (SNCs) of 98 ± 7 nm in size were commercially purchased from nanoCopisix and spin coated on the substrate. We optimized the spin coating process to increase the chance of getting the SNCs coupled to SPEs in the hBN transferred to Au/Si substrate.[42] Before spin coating, we sonicated the SNC solution for 5 minutes at room temperature to prevent clustering of the nanocubes and make a uniform solution. Then we placed 2 μL of SNC solution at the center of the Au/Si sample on the spin coater. We set it to spin 300 rpm for 50 sec, followed by 3000 rpm for 10 seconds. We verified the distribution of the SNCs on the hBN flake by optical microscopy before (Figure 2a) and after (Figure S7a) spin coating. We performed AFM imaging (Figure S7b) of a selected region along the hBN flake and confirmed the distribution and height (~ 100 nm) of the SNCs. Then, we performed fluorescence imaging of the targeted hBN flake with SNCs and found a high fluorescence intensity up to 5 Mc/s.[42] To further reduce further the fluorescence intensity (< 0.6 Mc/s) we used band pass filter in two wavelength bands of 600 – 650 nm and 700 – 750 nm. The fluorescence intensity map of a selected region with and without hBN flake showing both SPEs and SNCs is depicted in Figure S7c. The dashed line shows the borders/edges of the hBN flake, proving that there is autofluorescence from the nanocubes even without the hBN flake below.

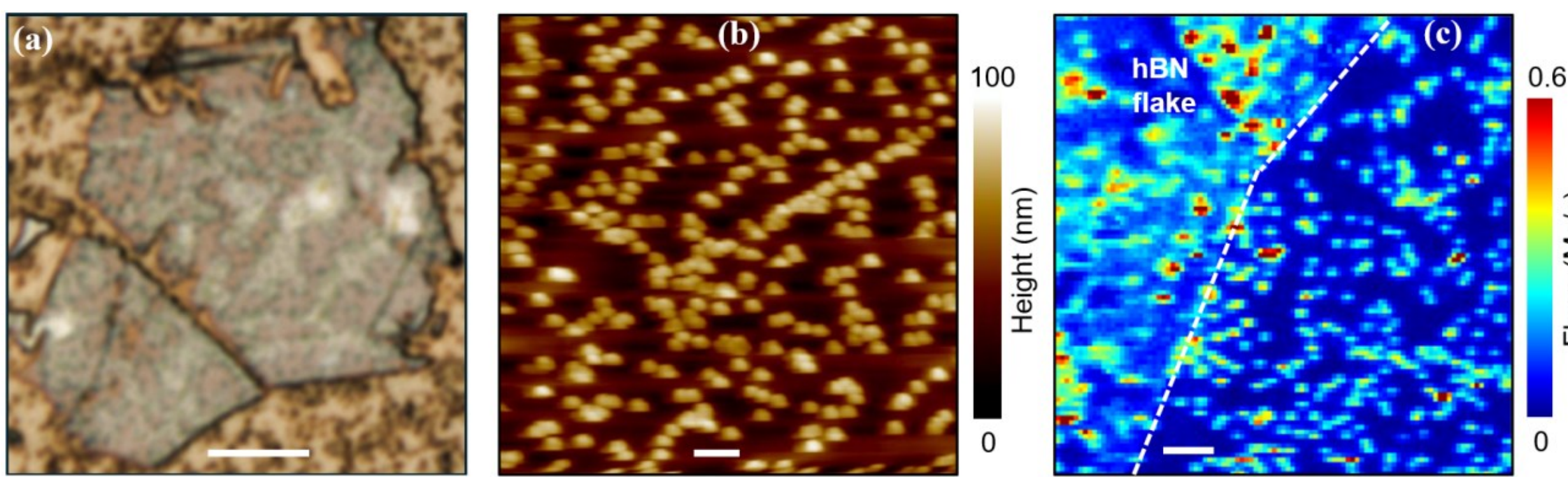

**Figure S7**. (a) Optical image of the flake after spin coating of 98-nm SNCs on top of the hBN flake. The scale bar in is 5 μm. (b) AFM image of a selected region along the hBN flakes, showing the distribution and height of the SNCs (the scale bar is 1 μm). (c) Fluorescence intensity map of a selected region with and without hBN flake showing both SPEs and SNCs (the scale bar is 2 μm). The dashed line shows the borders/edges of the hBN flake, proving that there is autofluorescence from the nanocubes even without the hBN flake below.

## S6. Optical and quantum properties of SPEs with metallic nanostructures

We measured the quantum optical properties of 68 SPEs before spreading the 98 nm SNCs. In Table S2, we summarize the optical properties (fluorescence spectrum, $g^{(2)}$ curve, lifetime curve, and fluorescence intensity vs excitation laser power saturation curve) of selected SPEs: SPE30, SPE34, SPE59, and SPE60 at room temperature. We further discuss the enhancement of these properties by adding SNCs in the main text and the following Sections S7 and S8. In Table S2 we show the optical properties (fluorescence spectrum, $g^{(2)}$ curve, lifetime curve, and fluorescence intensity vs excitation laser power saturation curve) of selected four SPEs from hBN/Au/Si substrate, labeled with the numbers SPE30, SPE34, SPE59, and SPE62, where 50% of the emitters show enhanced properties of SPEs in the presence of SNCs due to the plasmonic nanocavity effects discussed in the main text and below in Sections S7 and S8.

**Table S2**

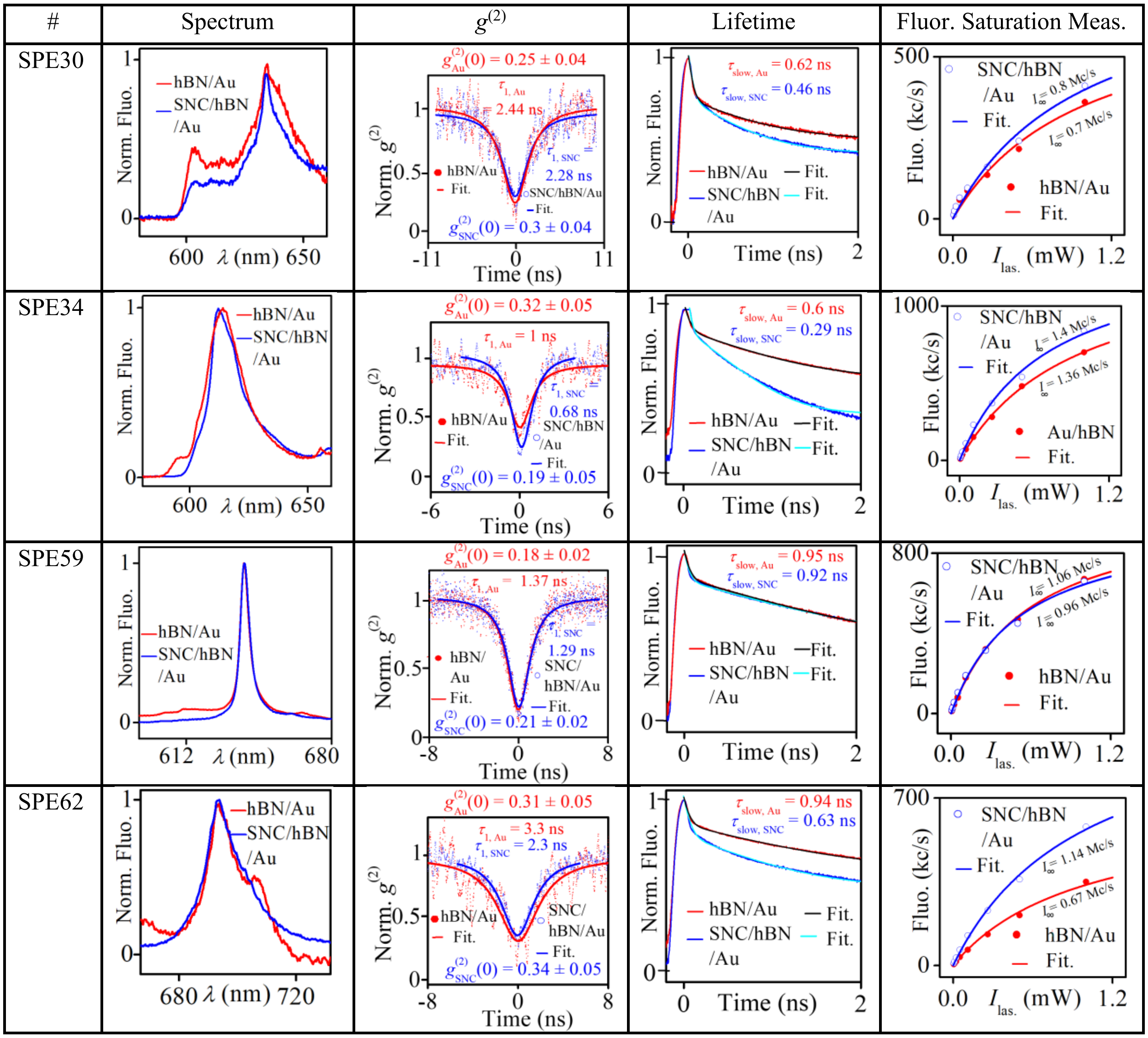

In Figure S8, we plot the histogram of the $g^{(2)}(0)$ dip values of the 68 SPEs studied in the hBN flake transferred to Au/Si substrate in Figures S1b and S1e. ~78% of the SPEs have $g^{(2)}(0)$ dip $\leq$ 0.4. The slight increase of $g^{(2)}(0)$ dip values in comparison to the SPEs in $SiO_2$/Si (Figure S5) is explained by the increased fluorescence background from the Au film.

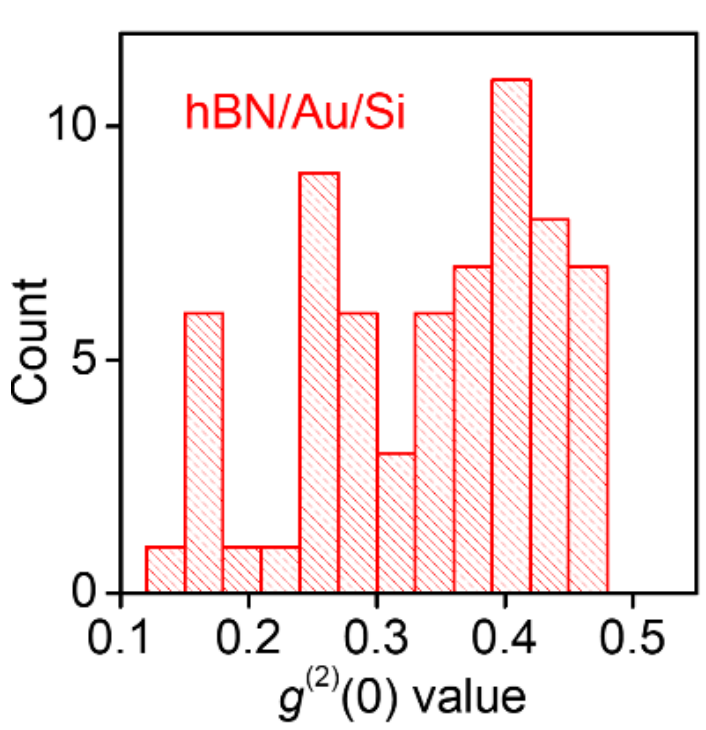

**Figure S8.** Histogram of $g^{(2)}(0)$ values of 68 SPEs studied on hBN/Au/Si substrate.

Moreover, in Figure S9, we plot the histograms of fluorescence saturation power ($I_\infty$) for SPEs in hBN/$SiO_2$, hBN/Au, and SNC/hBN/Au substrates, respectively. We extract a mean $I_{\infty,\mathrm{m}}$ value of 175 kc/s in hBN/$SiO_2$, 1.06 Mc/s in hBN/Au/Si, and 1.5 Mc/s in SNC/hBN/Au/Si. The overall enhancement factor of the plasmonic nanocavity is $\frac{I_{\infty,SNC/hBN/Au}}{I_{\infty,SiO2}} = \frac{1.5\ Mc/s}{0.214\ Mc/s} = 7$, corresponding to a fluorescence enhancement of 700%. The enhancement effect (increase of fluorescence) from just Au film is explained by the local confinement of electric field along the Au nanograins (see Figure S3d) due to excited plasmonic hotspots,[60,61] which is confirmed by theoretical simulations (see Section 2.4 and Section S7).

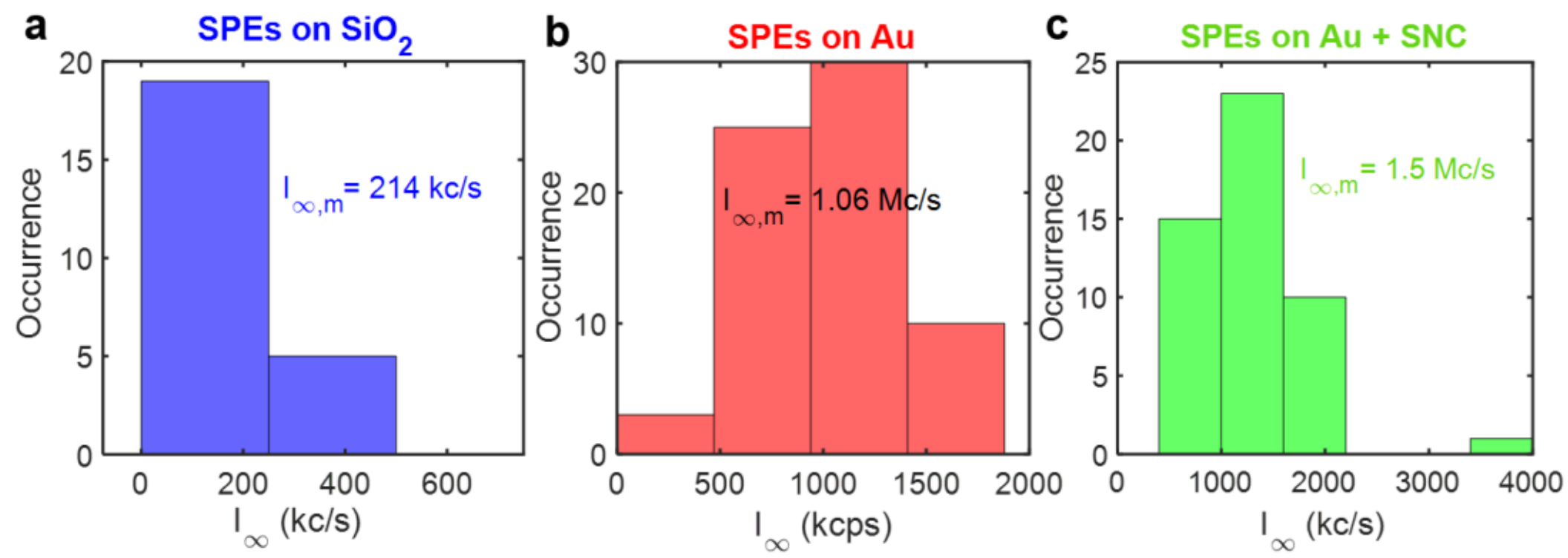

**Figure S9.** Histograms of fluorescence saturation power ($I_\infty$) for SPEs in (a) hBN/$SiO_2$, (b) hBN/Au, and (c) SNC/hBN/Au substrates, respectively.

### S7. Simulation results

The theoretical investigations of both scattering and emissivity properties were performed using the radio-frequency module of COMSOL Multiphysics. Further details are provided in the following subsections.

### S7.1. Geometrical parameters

To account for the roughness of the Au thin film fabricated on the Si substrate, we introduced hemi-ellipsoids on the metallic surface. These hemi-ellipsoids are spaced 5 nm apart from each other and have an in-plane base radius of 64 nm and an out-of-plane height radius of 14 nm. The radii values were determined based on both the average grain size and roughness assessments derived from the AFM images shown in Figure S3. Note the same roughness Au thin film is used for the simulations without and with the nanocube, as demonstrated in Figure 5 in the main paper. The thickness of the Au film is 103±1 nm obtained from AFM measurements (see Figure 1d). In our simulations, we considered a 10 nm curvature for both the corners and each side of the SNC, while maintaining 98 nm side length for the nanocube itself.

### S7.2. Dielectric functions

For the systematic theoretical characterization of both the scattering and emission properties of the proposed structure designs, the simulations also take into account the frequency-dependent complex dielectric functions ($\varepsilon$) of Si,[83] $SiO_2$,[84] and Ag[85] materials. The uniaxial dielectric nature of the multilayer hBN flakes were also incorporated into the simulations by considering the full dielectric tensor to be written as follows:[86,87]

$$\varepsilon_{hBN} = \begin{pmatrix} \varepsilon_{\perp} & 0 & 0 \\ 0 & \varepsilon_{\perp} & 0 \\ 0 & 0 & \varepsilon_{\parallel} \end{pmatrix},$$

where $\varepsilon_{\perp}$ is the relative dielectric constant perpendicular to the optic axis (i.e., in-plane), and $\varepsilon_{\parallel}$ is the relative dielectric constant parallel to the optic axis (i.e., out-of-plane).

### S7.3. Computed scattering cross section

To compute the scattering cross section, we employed the scattered-field formulation, which utilizes the analytical solution of the incident plane wave as the background electric field in the absence of SNC. The structure is enclosed by a spherical domain with a radius of 500 nm. In order to replicate the open space above the scatterer, scattering boundary conditions are applied to the outer shell of the spherical domain. The incident plane wave is polarized as transverse-magnetic (TM) and strikes the nanocube at normal incident angle. The scattering energy is determined by measuring the power of the field that scatters from the cube. This energy is then divided by the geometrical cross section of the SNC to calculate the scattering cross section. Due to the symmetry of the nanocube structure, similar scattering results are achieved when using an incident plane wave that is transverse-electric (TE) polarized or one that is incoherently polarized and contains both TE and TM polarizations.

As it can be seen from Figure S10a, in the absence of SNC, a single broad (550 – 700 nm) resonance mode is observed in the computed spectrum of scattering cross section which red shifts as a function of the hBN layer thickness. In the presence of SNC (Figure S10b), a more narrowband resonance mode (700 – 750 nm) emerges in the spectral map. This is due to the creation of a plasmonic nanocavity, where the electric field enhancement is boosted inside hBN compared to the case when there is no cavity, as demonstrated in Figure 5b and Figure 5e, respectively.[12,43]

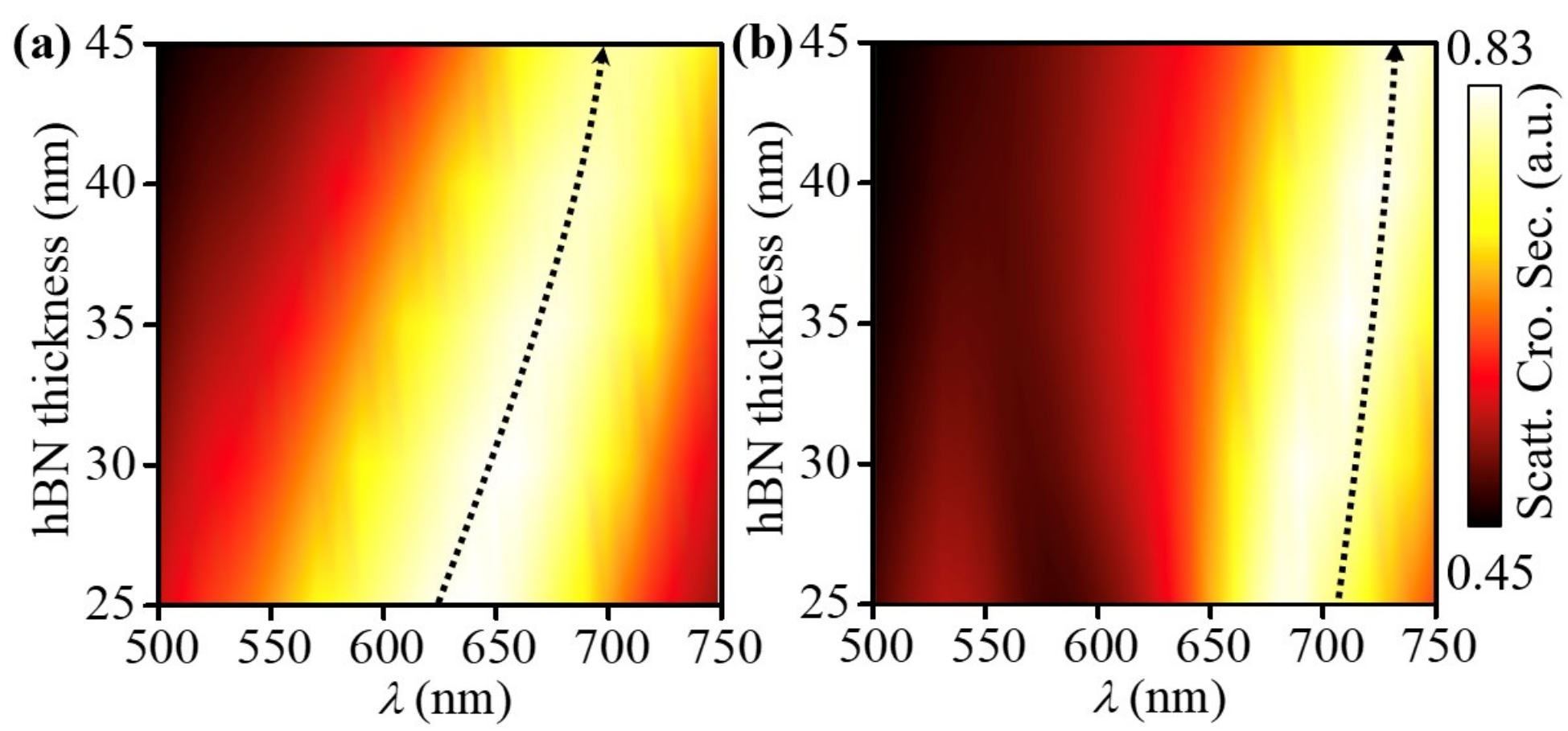

**Figure S10.** Color density plot of scattering cross section spectra as a function of hBN layer thickness for the proposed systems: (a) without and (b) with 98-nm SNC.

### S7.4. Theoretical investigation of emission properties

As previously demonstrated by several studies, the plasmonic nanocavity results in strong field enhancement due to the localized gap plasmon that can be utilized to engineer the spontaneous emission rate.[12,43,88] However, it was also observed that emission systems with plasmonic inclusions suffer from elevated nonradiative decay and ohmic losses leading to quenching.[88–90]

This trade-off requires a careful systematic theoretical analysis to find out the optimum ultrafast photon emission process of the proposed metallic nanocavity systems.

### S7.4.1. Approach to extract emission properties

Hence, similar to the scattering cross section characterization, simulations to compute the emission properties are performed in the spherical domain mentioned in the previous section. To compute the radiative and nonradiative decay rates and quantum efficiency parameters, a monochromatic point dipole is employed that is used to obtain the spatial distribution and spectral evolutions of the aforementioned parameters (see Figure 5, Figure 6, and Figure S11). The computation of the Green's function of the system is performed based on the monochromatic point dipole emitter approach.[62] The dipole emitter is swept on a discrete 20 by 20 grid system. This grid system is placed at exact spacings for both rough Au thin film surface without nanocube and the presence of SNC on the rough Au film. The array of point dipoles is also placed at different heights (along the *z* axis) inside the hBN layer to extract the depth profile of emission properties (see Figure S11). The integration of the total power radiated out of the entire domain and absorbed from the metallic inclusions of the system forms the main basis of the radiative and nonradiative decay rates calculations.

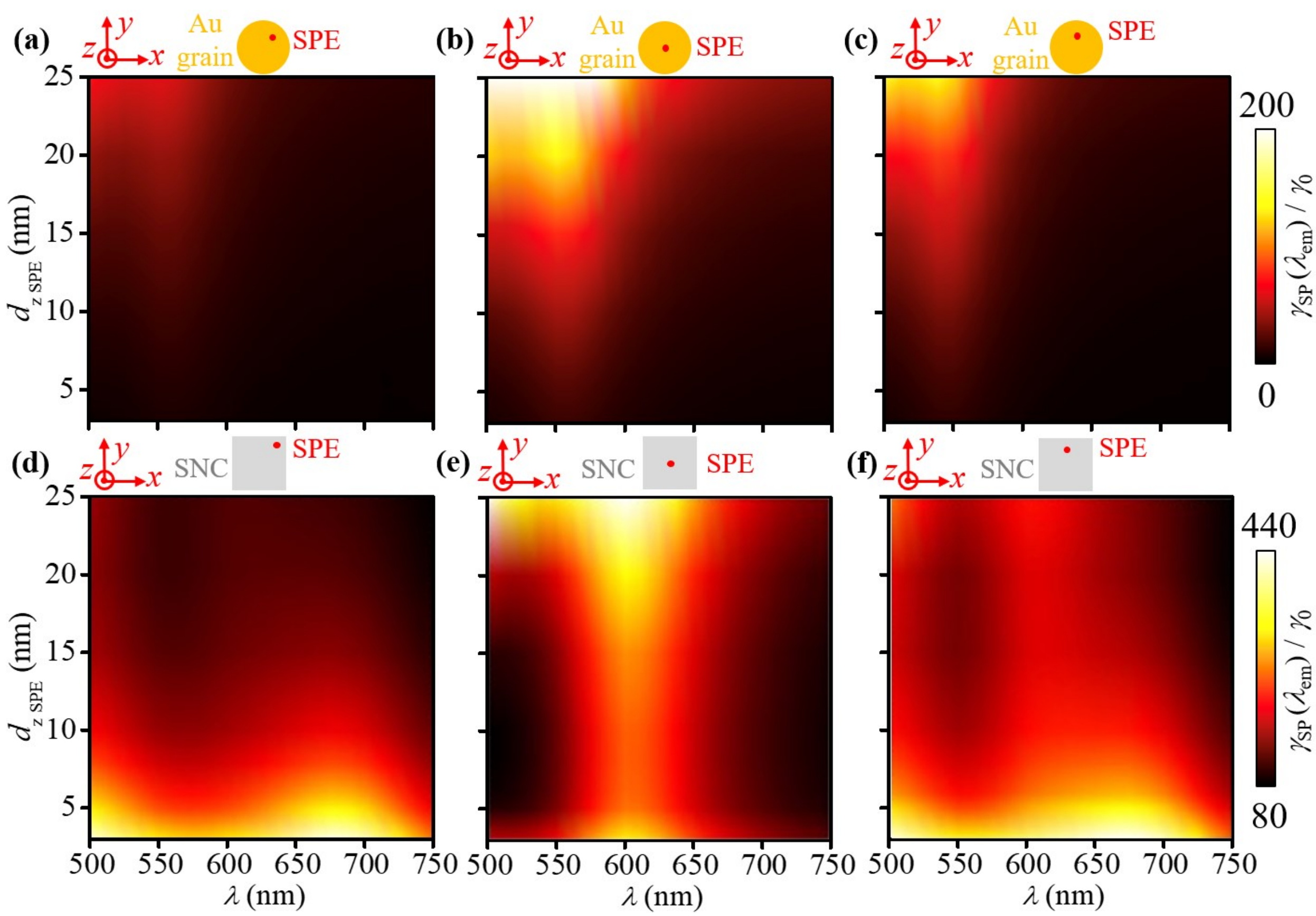

**Figure S11.** Color density plot of hBN layer thickness dependent spectral evolution of the normalized spontaneous emission rate ($\gamma_{sp}/\gamma_0$) for the nanostructure without (a, b, and c) and with (d, e, and f) SNC for different emitter locations which are schematically illustrated on top of each figure.

### S7.4.2. Quantum Yield

From the emission perspective, the quantum yield (QY) plays a fundamental role as it is the probability of having a photon radiated during a relaxation process given by $QY = \gamma_r/\gamma_{sp}$, where $\gamma_r$ is the radiative decay rate and $\gamma_{sp}$ is the spontaneous emission rate.[62] The spontaneous emission spectrum of the emitter designs with and without SNC are presented as a function of hBN thickness in Figure S11. Three different locations are selected to present the spontaneous emission response of our designs: at the center, in the middle of one side, and at one corner of SNC. As a relevant comparison, we selected the same locations for the design without SNC. For the latter case, the roughness of Au thin film was represented as the hemi-ellipsoids on the surface (the geometrical details are provided in Figure 5, Figure 6, and the Supporting Information Sections S2 and S7.1). We observe that when SPE is close to the bottom surface of SNC, we obtain a much larger spontaneous emission rate as the plasmonic nanocavity localizes the electric field in the close vicinity of SNC. However, for the emitter design without SNC, we do not create a nanocavity, instead, the hemi-ellipsoidal roughness inclusions result in localized plasmonic hotspots that lead to higher spontaneous emission rate when the SPE is close to the ellipsoidal inclusions (SPE $z$ height: $d_{z,SPE} > 15$ nm) beneath the hBN layer.

### S8. Metallic nanocavity enhancement of quantum properties of emitter SPE26 in hBN/Au/Si

We focus here on the enhanced quantum properties of selected emitters in the hBN flakes deposited on $SiO_2$ (e.g., SPE23 highlighted in Figure S1b) and Au/Si (e.g., SPE26 highlighted in Figure S12a) emitting at a wavelength of 750 nm with and without SNCs. The metallic plasmonic nanocavity leads to a decrease in the FWHM emission by 50% (inset of Figure S12b) and a significant narrowing of the $g^{(2)}$ response (Figure S12b) with a decrease of $\tau_1$ from 3.34 ± 0.07 ns in hBN/$SiO_2$ to 1.2 ± 0.05 ns in hBN/Au/Si without SNC, and to 0.41 ± 0.02 ns in hBN/Au/Si with SNC, respectively. $g^{(2)}(0)$ is 0.11 ± 0.01 for SPE23 in hBN/$SiO_2$, 0.36 ± 0.02 for SPE26 in hBN/Au without SNC, and 0.46 ± 0.02 in hBN/Au with SNC. In Figure S12c, we plot the lifetime curves of emitters SPE23 in hBN/$SiO_2$ and SPE26 in hBN/Au without and with SNC. The measured curves of SPEs in hBN/$SiO_2$ are well fitted with one exponential decay function $a_3 e^{-t/\tau_{3,SiO_2}}$, where $t$ is the time, $\tau_{3,SiO_2}$ is the emission lifetime, and $a_3$ a weighing factor.[30] $\tau_{3,SiO_2}$ is 4.13 ± 0.16 ns for SPE23 in hBN/$SiO_2$. The measured lifetime curves in hBN/Au without and with SNC are fitted (solid lines) with two exponential decay functions: $a_{fast} e^{-t/\tau_{fast}} + a_{slow} e^{-t/\tau_{slow}}$, where $a_{fast}$ and $a_{slow}$ are weighing factors, where $\tau_{slow}$ is 1.085 ns and 0.3 ns for SPE26 in hBN/Au without and with SNC, respectively, while $\tau_{fast}$ is 41 ps and 19 ps for SPE26 in hBN/Au without and with SNC, respectively.

The plasmon SPE enhancement rate from the Au film in comparison to the $SiO_2$ substrate is: $\frac{1/\langle\tau_{fast\,Au}\rangle}{1/\langle\tau_{3,\mathrm{SiO_2}}\rangle} = \frac{\langle\tau_{3,\mathrm{SiO_2}}\rangle}{\langle\tau_{fast\,Au}\rangle} = \frac{4.13\,ns}{0.041\,ns} = 100$, corresponding to an enhancement factor of 100 times. We define the plasmonic nanocavity SPE enhancement in hBN/Au with and without SNC as: $\frac{1/\langle\tau_{fast\,\mathrm{Au\,w.\,SNC}}\rangle}{1/\langle\tau_{fast\,\mathrm{Au\,w/o\,SNC}}\rangle} = \frac{\langle\tau_{fast\,\mathrm{Au\,w/o\,SNC}}\rangle}{\langle\tau_{fast\,\mathrm{Au\,w.\,SNC}}\rangle} = \frac{44\,ps}{19\,ps} = 2.13$, corresponding to an additional enhancement rate of 231%. The overall enhancement of fluorescence (spontaneous emission) lifetime from SPE23 in hBN/$SiO_2$ in comparison to SPE26 in SNC/hBN/Au nanocavity is 213 times.

To further illustrate the effect of the plasmonic nanocavity in the fluorescence intensity of the SPEs, we measure their saturated count rate $I_\infty$ in hBN/$SiO_2$, and hBN/Au before and after spreading the SNCs. In Figure S12d, we plot the measured (scattered curves) fluorescence intensity of SPE23 in $SiO_2$ and SPE26 in hBN/Au (without and with SNCs) as a function of the CW green laser power. The saturation curves are fitted (solid lines in Figure S12d) to:[41,42] $I = \frac{I_\infty P}{(P_{sat}+P)}$, where $P_{Sat}$ is the saturation power. $I_{\infty,SiO2}$ = 0.24 Mc/s, $I_{\infty,hBN/Au}$ = 0.64 Mc/s, and $I_{\infty,SNC/hBN/Au}$ = 1.09 Mc/s for hBN/$SiO_2$, hBN/Au, and SNC/hBN/Au, respectively. The overall enhancement factor of the plasmonic nanocavity is $\frac{I_{\infty,SNC/hBN/Au}}{I_{\infty,SiO2}} = \frac{1.09\ Mc/s}{0.24\ Mc/s}$ = 4.54, corresponding to a fluorescence enhancement of 454%.

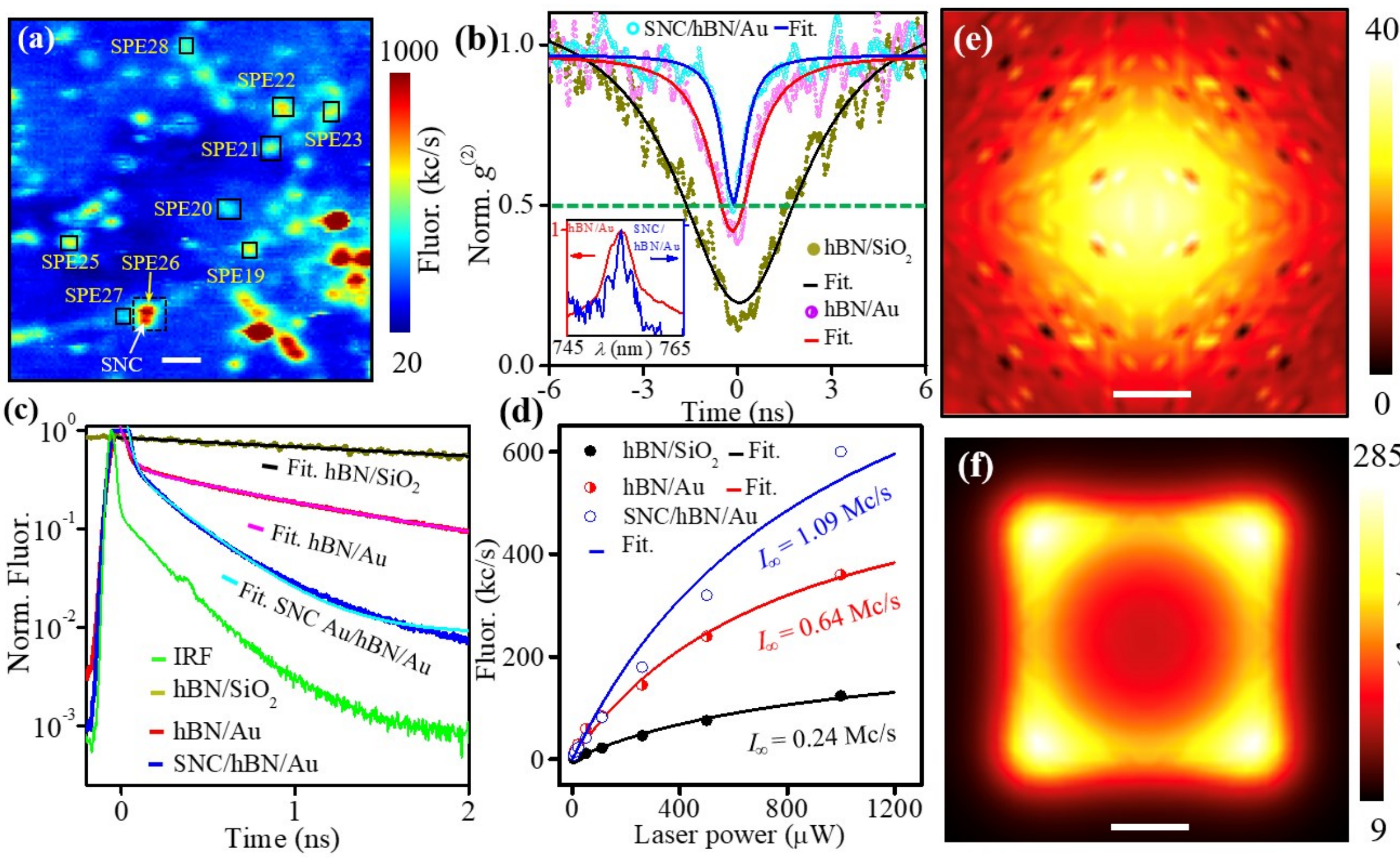

**Figure S12.** (a) Fluorescence map of a region of the hBN flake with SNCs shown in Figure 1e. The scale bar is 2 μm. Some of the SPEs are coupled with SNCs such as emitter SPE26. Insert of (b): Measured spectrum of SPE26 before and after depositing the 98 nm SNCs. (b) Measured autocorrelation $g^{(2)}$ intensity as a function of time for a selected SPE in hBN flake transferred to $SiO_2$ (filled circles) and for SPE26 (highlighted with dashed squares in (a) in the hBN flake transferred to Au/Si substrate without (half-filled circles) and with (open circles) SNCs. The measured curves are fitted (solid lines) to the function $g^{(2)}(t) \simeq 1-(1+a_1)e^{-t/\tau_1}$. (c) Measured lifetime response as a function of time for a SPE23 in hBN flake transferred to $SiO_2$ (solid dark yellow line) and for SPE26 in the hBN flake transferred to Au/Si substrate without (solid purple line) and with (solid aqua line) SNCs. The measured curves are fitted (solid lines) to $a_2e^{-t/\tau_{fast}} + a_3e^{-t/\tau_{slow}}$. The instrument response function (IRF) of the setup is plotted in solid green light and exhibits a time response on the order of 19 ps. (d) Measured (scattered) and calculated (solid lines) fluorescence intensity versus CW green laser power for SPE23 in hBN flake transferred to $SiO_2$ (filled circles) and for SPE26 highlighted in (a). The computed spatial distribution of normalized spontaneous emission rate ($\gamma_{sp}(d_{z,SPE} = 5\text{ nm})/\gamma_0$) for the SPE emitter design without (e) and with (f) SNC at the emission wavelength of 754 nm. The SPE $z$ position is depicted in Figure 5a and Figure 5d with dashed lines. The scale bars in (e) and (f) are equal to 20 nm.

The corresponding spatial distribution of normalized spontaneous emission rate at the emission wavelength of 754 nm for emitter (SPE26) designs without and with SNC are theoretically computed from simulations and are presented in Figures S12e and S12f, respectively. For these simulations, the SPE *z* height ($d_{z,SPE}$) is chosen as 5 nm. For the emitter design without SNC, the normalized spontaneous rate is more pronounced at the center of the system which can be attributed to the hemi-ellipsoidal roughness inclusions that touch to the hBN layer only at center of the presented spatial distribution (see Figure S12e). Upon incorporating the SNC into the emitter system design, the emission rate experiences a significant enhancement, surpassing 100 times that of the design without SNC. As depicted in Figure S12f, analogous to the findings in Figure 5f, the emission intensifies at the corners of the SNC. This strong emission characteristics are directly linked to the resulted plasmonic nanocavity, which facilitates the localization of the electromagnetic field at the bottom corners and edges of the SNC.